\documentclass[twocolumn]{aastex631}
\usepackage[utf8]{inputenc}

\usepackage[nointegrals]{wasysym}
\usepackage{amsmath}
\usepackage{graphicx}
\usepackage{commath}
\usepackage{bm}
\usepackage{float}

\newcommand{\HST}{\textit{HST }}
\newcommand{\TESS}{\textit{TESS }}
\newcommand{\Gaia}{\textit{Gaia }}
\newcommand{\angstrom}{\textup{\AA}}

\begin{document}

\begin{abstract}
     Interpreting the short-timescale variability of the accreting, young, low-mass stars known as Classical T Tauri stars remains an open task. Month-long, continuous light curves from the Transiting Exoplanet Survey Satellite (\textit{TESS}) have become available for hundreds of T Tauri stars. With this vast data set, identifying connections between the variability observed by \TESS and short-timescale accretion variability is valuable for characterizing the accretion process. To this end, we obtained short-cadence \TESS observations of 14 T Tauri stars in the Taurus star-formation region along with simultaneous ground-based, UBVRI-band photometry to be used as accretion diagnostics. In addition, we combine our dataset with previously published simultaneous NUV-NIR \textit{Hubble Space Telescope} spectra for one member of the sample. We find evidence that much of the short-timescale variability observed in the \TESS light curves can be attributed to changes in the accretion rate, but note significant scatter between separate nights and objects. We identify hints of time lags within our dataset that increase at shorter wavelengths which we suggest may be evidence of longitudinal density stratification of the accretion column. Our results highlight that contemporaneous, multi-wavelength observations remain critical for providing context for the observed variability of these stars.
\end{abstract}

\title{Understanding Accretion Variability Through TESS Observations of Taurus}

\author[0000-0003-1639-510X]{Connor E. Robinson}
\affiliation{Department of Physics \& Astronomy, Amherst College, C025 Science Center 25 East Drive, Amherst, MA 01002, USA}
\affiliation{Department of Astronomy \& Institute for Astrophysical Research, Boston University, 725 Commonwealth Avenue, Boston, MA 02215, USA}
\author[0000-0001-9227-5949]{Catherine C. Espaillat}
\affiliation{Department of Astronomy \& Institute for Astrophysical Research, Boston University, 725 Commonwealth Avenue, Boston, MA 02215, USA}
\author[0000-0001-8812-0565]{Joseph E. Rodriguez}
\affiliation{Center for Data Intensive and Time Domain Astronomy, Department of Physics and Astronomy, Michigan State University, East Lansing, MI 48824, USA}

\correspondingauthor{Connor Robinson}
\email{corobinson@amherst.edu}

\section{Introduction \label{s:introduction}}

Classical T Tauri stars (CTTS) are young, low-mass stars that are accreting material from the protoplanetary disk that forms and evolves alongside the star. Accretion is a critical part of the star--disk evolution process because it produces much of the high-energy radiation field that permeates the inner regions of the disk. Accretion also provides a method of probing the inner regions of the disk, which traditional observational techniques lack the ability to resolve. For two recent reviews on accretion onto CTTS, see \citet{hartmann16} and \citet{schneider20}.

Nearly all CTTS show some manner of variability. Sources of the variability include rotational modulation from hot/cool spots, changes in the mass accretion rate, stellar activity, and variable extinction from material in the surrounding disk or dust entrained within the stellar magnetosphere \citep[e.g.,][]{herbst94, cody14}. Simulations of the inner regions of disks have been able to explain some of the observed variability caused by changes in the mass accretion rate on hour- and day-timescales via instabilities in the inner regions of the disk and overdensities in the accretion column powered by turbulence \citep[e.g.,][]{kurosawa13, robinson21}. 

Despite the above progress, variability on minute- to hour- timescales and the driving forces behind it remain relatively unclear. There are only a handful of studies that exist with long-term, short-cadence (defined here to be 1- to 2-minute) light curves of accreting young stars \citep[e.g.,][]{siwak18} . The Transiting Exoplanet Survey Satellite \citep[\textit{TESS},][]{ricker15} has begun to provide near-continuous, high-precision light curves of CTTS which can probe variability that may have been missed by previous studies \citep[e.g.,][]{zsidi22}.
However, it is currently unclear how much of the variability as observed through the \TESS bandpass can be attributed to accretion, and how much is driven by other sources of variability. Sets of simultaneous measurements of the accretion rate are thus critical when interpreting \TESS light curves for studies of accretion.

Ground-based photometry can be used to measure the accretion rate by relating the excess emission above photospheric levels to the total accretion luminosities using empirical bolometric corrections \citep[e.g.,][]{gullbring98}. The accretion luminosity can then be transformed into an estimate of the accretion rate using the stellar mass and radius from stellar evolutionary models \citep[e.g.,][]{baraffe98}, and by assuming that the material is falling at the free fall velocity under the magnetospheric accretion paradigm. 

Space-based UV spectra can also be used to measure the accretion rate. In particular, contemporaneous FUV - NIR Hubble Space Telescope (\textit{HST}) observations with the Space Telescope Imaging Spectrograph (STIS) have been a powerful tool in previous studies for characterizing the accretion shock \citep[e.g.,][]{ingleby11, ingleby13, ingleby15, robinson19}. The accretion shock models of \citet{calvet98} predict that the shape of the UV continuum excess is a function of the kinetic energy of the accretion flow. Under the assumption of the flow being in free fall, the kinetic energy flux is proportional to the density of the accretion column. Thus, the shape of the UV continuum can be used to break the degeneracy between accretion column density and surface coverage that is present for optical measurements of the accretion rate \citep[e.g.,][]{ingleby13, ingleby15, robinson19}.

To test the connection between the variability in TTS as observed by \TESS to accretion, we present short-cadence \TESS observations and simultaneous ground-based UBVRI photometry of 14 TTS in the Taurus star forming region. In addition, we compare our results to five previously published simultaneous HST NUV-NIR STIS spectra and present one new epoch for one member of the survey, GM Aur \citep[see ][]{espaillat21}. This object is a young K5 star with a transitional disk containing residual optically thin dust in the cavity \citep{calvet05, hughes09, andrews11, espaillat11, macias18} and has been the focus of several other recent \HST UV studies of accretion studies of accretion \citep{ingleby15, robinson19}.  GM Aur is an ideal target for studies of accretion because of its robust accretion rate, similarity to the young Sun, inner cavity, and moderate inclination of $52^\circ.77^{+0.05}_{-0.04}$ \citep{macias18}, which limits the chance for disk occultation of the star. In \S\ref{s:observations}, information about the three data sets and reduction processes are presented. In \S\ref{s:analysis}, we detail our modeling efforts and the steps taken to calculate accretion rates. In \S\ref{s:results}, we present the results of our analysis and we discuss the significance of these results in \S\ref{s:discussion}. We summarize our key findings in \S\ref{s:summary}.

\section{Observations \label{s:observations}}
\subsection{\textit{TESS}}

\TESS offers the opportunity to obtain continuous, (aside from a short gap due to downlinking at orbital perigee) month-long light curves for many TTS. The \TESS spectral response curve is very broad, and covers roughly between $0.6 \, \rm{\mu m}$ and $1.1 \, \rm{\mu m}$ with a steep cutoff at the blue end. While the peak of excess flux from accretion is in the NUV, excess continuum emission extends into the optical and near IR where \TESS is sensitive \citep{calvet98}. Short-cadence (2-minute) light curves of the 14 targets included in our sample were obtained in the \TESS GO program 22216 (PI: C. Robinson) as a part of the \TESS Sector 19 observing campaign (Nov. 28th - Dec 23rd, 2019).
At the time that our objects were observed, two types of data products were available from \textit{TESS}: Full Frame Images (FFI) and short-cadence observations. FFIs are wide-FOV images that cover a $24^\circ \times 90^\circ$ strip of the sky (defined as a \TESS data sector) binned to a cadence of 30 minutes aboard the spacecraft. Short-cadence observations are produced for specific individual stars and the data is binned to a 2 minute cadence.  Our sample was formed by comparing the population of T Tauri stars from \citet{andrews13} with the \TESS Sector 19 footprint and selecting from the remaining members those that we could recover at sufficient SNR in the short-cadence mode of \TESS. None of these targets have been previously observed by K2, making these the first short-cadence, long-baseline observations of these targets. The \TESS magnitudes for our sample range between 12 and 9. SU Aur was originally included in our sample, but because it landed on the edge of the \TESS detector, we exclude it from our analysis.

To verify the quality of our 2-minute cadence observations, the 30-minute FFI were inspected using the Python package \texttt{Lightkurve} \citep{lightkurve18}. Transient (presumably solar-system) objects occasionally cross through the field of view, but from visual inspection, these transient objects do not significantly impact the quality of the light curves. Because of the large size of the \TESS pixels ($21''$) contamination can be significant in crowded regions. From visual inspection of previous ground-based images of the sample, we note the stars in our sample are significantly brighter than any secondary sources within the extraction region. To further probe the degree of contamination from nearby stars, we applied a variety of test apertures to the FFI to produce light curves for comparison against the short-cadence light curve produced by the Science Processing Operations Center (SPOC) pipeline \citep{smith12, stumpe14, jenkins16}. From this analysis, we found only minor contamination from neighboring stars and that light curves from the FFI with these apertures closely match the short-cadence light curve produced by the SPOC pipeline. Given this, we adopt the SPOC pipeline reduction. We also monitored the background flux in the FFI to determine the length of time around the orbit gaps to mask and found the standard SPOC mask is suitable to avoid contamination. Erroneous observations (e.g., single-frame sudden brightening/dimming events) were identified by visual inspection and masked. The light curves can be seen in Figure~\ref{fig:TESS_LC}. Comments on the \TESS light curves for each object are included in the appendix (Section \ref{s:indiv}).

\begin{figure*} 
    \centering
    \includegraphics{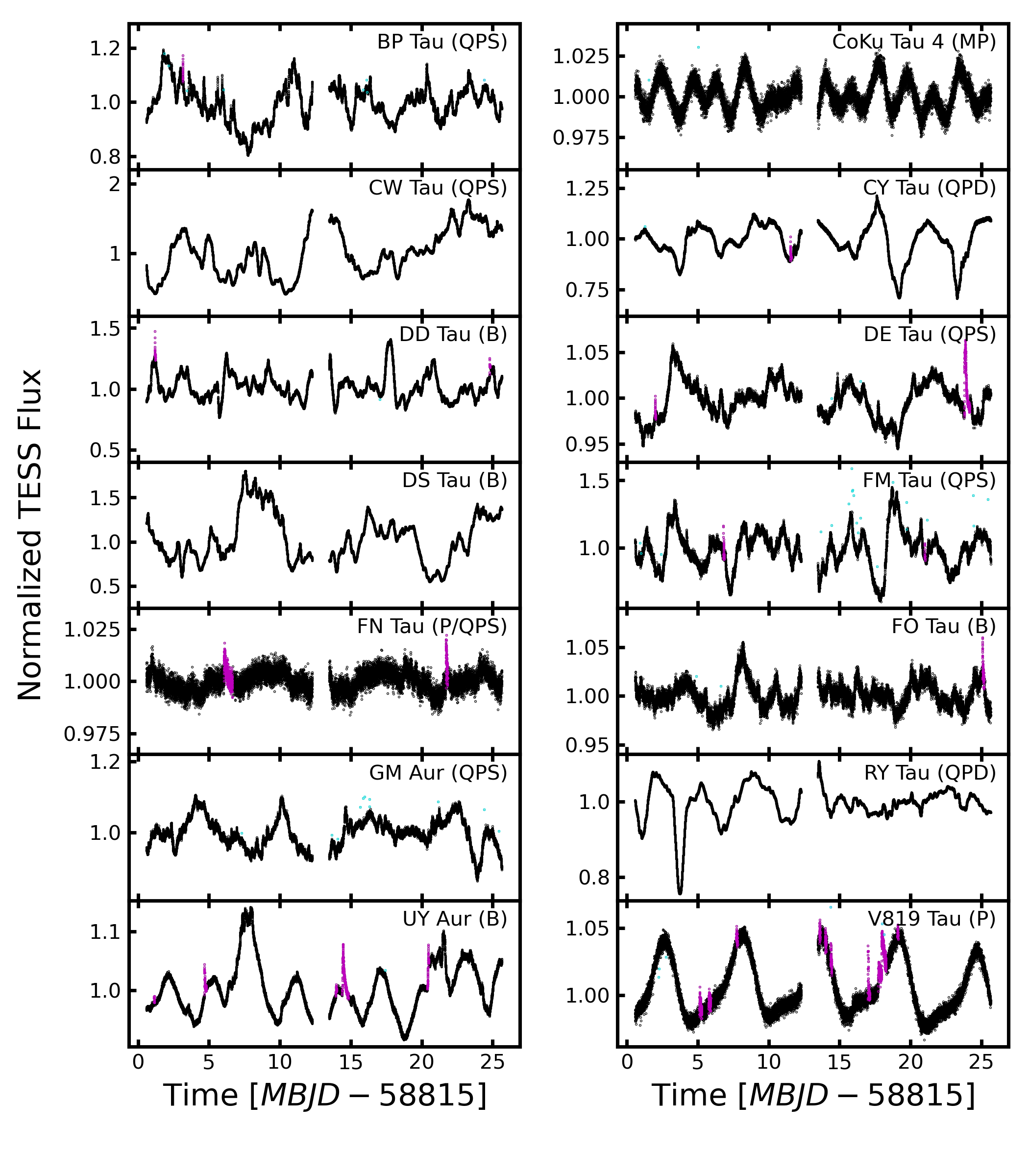}
    \caption{Short-cadence (2 minute) \TESS light curves for the 14 objects in the sample. The photometry shown here was produced using the standard SPOC pipeline. The timestamps of the data are presented as Modified Barycentric Julian Dates (MBJD) in the Barycentric Dynamical Time (TDB) scale. 
    Flares identified by-eye by their characteristic rapid rise and exponential decay morphology have been flagged in magenta, and single-point outliers have been flagged in cyan. The light curves are shown normalized by their median value (this median includes the flagged points). We exclude these flagged points in the time-lag analysis presented in \S~\ref{sss:cross_ref_lag}. The light curve classification for each object is listed in parentheses to the right of the object's name (see \S~\ref{ss:QM} and Table~\ref{tab:QMT}).}
    \label{fig:TESS_LC}
\end{figure*}

\subsection{LDT \label{ss:ldt_obs}}
We obtained UBVRI optical photometry for each object using the Large Monolithic Imager (LMI) on the 4.3m Lowell Discovery Telescope (LDT) in Happy Jack, AZ. A set of UBVRI images were obtained for each target during each of six nights spread over Dec. 2019 (specifically 2019-12-02, 2019-12-07, 2019-12-10, 2019-12-13, 2019-12-18, and 2019-12-21 UT). These data were taken simultaneously with the \TESS observations (with the exception of some of the observations on 2019-12-10 which occurred near the start of the \TESS downlink gap). Exposure times for these observations varied depending on object brightness and weather conditions. Exposure times for B, V, R and I were in the range of 0.1 to 10 s and between 10 and 120 s for U. In addition, we obtained six 1-2 hr monitoring series of $\sim 30$ s U-band exposures of GM Aur during the same nights.  Standard flat fielding and bias subtraction were applied to the images before extraction of photometry. Differential photometry was performed using the ensemble method for inhomogeneous sets of exposures from \citet{honeycutt92} using a custom \texttt{Jupyter} widget\footnote{We designed our differential photometry widget to be flexible, interactive, and easy to use, making it a powerful tool with both research and education applications. Living open source code and instructions on its use can be found at \url{https://github.com/connorrobinson/ensemble}. A frozen release can be found here: \citet{robinson_ensemble22}.} and the \texttt{astropy} package \texttt{photutils} \citep{bradley20}. After our initial source identification process, we identify and exclude variable stars by searching for stars that lie above the locus of points when plotting mean instrumental magnitude against the standard deviation of the instrumental magnitude.  Depending on the population of background stars in the field, we used between 3 and 287 comparison stars to perform our differential photometry. Stars that only had a few ($< 10$) comparison stars, which may make their relative photometry slightly less reliable, include FM Tau (3), CY Tau (4), and CW Tau (4). For additional details on this method of inhomogeneous differential ensemble photometry, see \citet{honeycutt92}.

To transform our instrumental magnitudes to an absolute system, we measured the zero point using the standard field centered around GD 64 \citep{landolt13} on 2019-12-21 UT. That night was chosen in particular because our observations during that time have the smallest variance in atmospheric extinction. Uncertainty estimates for our instrumental magnitudes were calculated following the procedures of \citet{honeycutt92}, which includes both an estimate of the measurement uncertainty in the flux (arising from Poisson and read noise) and the uncertainty in the atmospheric extinction present in that that exposure. We then add this uncertainty in our instrumental magnitudes in quadrature with an uncertainty in the measurement uncertainty of our zero point (estimated through typical error propagation techniques) to obtain our final estimate of uncertainty in our photometry.  A table containing our UBVRI apparent magnitudes for the sample is presented in Table~\ref{tab:LDT_table}. We show the results from our GM Aur U-band monitoring series in Table~\ref{tab:gmaur_Uband}. 

\begin{deluxetable*}{cccccccccc}
\centering
\tabletypesize{\scriptsize}
\tablecolumns{10}
\tablecaption{LDT UBVRI photometry and $\dot{M}$ measurements for the entire sample\label{tab:LDT_table}}
\tablehead{\colhead{Name} & \colhead{Exposure Start} & \colhead{U} & \colhead{B} & \colhead{V} & \colhead{R} & \colhead{I} & \colhead{$\rm{\dot{M}_{fixed}}$} & \colhead{$\rm{\dot{M}_{rand}}$} & \colhead{$\rm{\dot{M}_{sys}}$} \vspace{-2mm} \\  & [MBJD] & [mag] & [mag] & [mag] & [mag] & [mag] & \colhead{$\mathrm{[10^{-8}M_\odot/yr]}$} &  \colhead{$\mathrm{[10^{-8}M_\odot/yr]}$} & \colhead{$\mathrm{[10^{-8}M_\odot/yr]}$} } 
\startdata
BP Tau & 58819.4580385 & $12.42^{+0.08}_{-0.08}$ & $12.612^{+0.009}_{-0.009}$ & $12.01^{+0.03}_{-0.03}$ & $11.198^{+0.026}_{-0.026}$ & $10.304^{+0.028}_{-0.028}$ & 2.2 & $2.2^{+0.3}_{-0.3}$ & $2.6^{+1.1}_{-0.8}$ \\
BP Tau & 58824.1239826 & $12.76^{+0.07}_{-0.07}$ & $13.055^{+0.009}_{-0.009}$ & $12.214^{+0.029}_{-0.029}$ & $11.409^{+0.025}_{-0.025}$ & $10.443^{+0.026}_{-0.026}$ & 1.61 & $1.61^{+0.26}_{-0.23}$ & $1.9^{+0.8}_{-0.6}$ \\
BP Tau & 58827.1278698 & $12.04^{+0.07}_{-0.07}$ & $12.403^{+0.009}_{-0.009}$ & $11.831^{+0.029}_{-0.029}$ & $11.203^{+0.025}_{-0.025}$ & $10.410^{+0.026}_{-0.026}$ & 3.1 & $3.1^{+0.5}_{-0.4}$ & $3.8^{+1.6}_{-1.2}$ \\
BP Tau & 58830.1149533 & $12.62^{+0.07}_{-0.07}$ & $12.972^{+0.009}_{-0.009}$ & $12.18^{+0.03}_{-0.03}$ & $11.393^{+0.025}_{-0.025}$ & $10.473^{+0.026}_{-0.026}$ & 1.83 & $1.83^{+0.28}_{-0.26}$ & $2.2^{+1.0}_{-0.7}$ \\
BP Tau & 58835.1078396 & $12.45^{+0.07}_{-0.07}$ & $12.768^{+0.009}_{-0.009}$ & $12.013^{+0.029}_{-0.029}$ & $11.290^{+0.025}_{-0.025}$ & $10.391^{+0.026}_{-0.026}$ & 2.1 & $2.1^{+0.3}_{-0.3}$ & $2.6^{+1.2}_{-0.8}$ \\
BP Tau & 58838.1136513 & $12.30^{+0.07}_{-0.07}$ & $12.607^{+0.009}_{-0.009}$ & $11.916^{+0.029}_{-0.029}$ & $11.213^{+0.025}_{-0.025}$ & $10.388^{+0.026}_{-0.026}$ & 2.5 & $2.4^{+0.4}_{-0.3}$ & $3.0^{+1.2}_{-0.9}$ \\
CoKu Tau 4 & 58819.3453997 & $18.04^{+0.07}_{-0.07}$ & $16.471^{+0.015}_{-0.015}$ & $14.72^{+0.03}_{-0.03}$ & $13.42^{+0.03}_{-0.03}$ & $11.92^{+0.03}_{-0.03}$ & 0.09 & $0.091^{+0.03}_{-0.020}$ & $0.10^{+0.10}_{-0.06}$ \\
CoKu Tau 4 & 58824.2051132 & $17.94^{+0.07}_{-0.07}$ & $16.470^{+0.015}_{-0.015}$ & $14.75^{+0.03}_{-0.03}$ & $13.43^{+0.03}_{-0.03}$ & $11.94^{+0.03}_{-0.03}$ & 0.11 & $0.115^{+0.03}_{-0.028}$ & $0.12^{+0.09}_{-0.07}$ \\
CoKu Tau 4 & 58827.2138518 & $18.01^{+0.07}_{-0.07}$ & $16.457^{+0.015}_{-0.015}$ & $14.73^{+0.03}_{-0.03}$ & $13.43^{+0.03}_{-0.03}$ & $11.92^{+0.03}_{-0.03}$ & 0.10 & $0.100^{+0.03}_{-0.025}$ & $0.10^{+0.09}_{-0.06}$ \\
CoKu Tau 4 & 58830.2006564 & -- & $16.513^{+0.016}_{-0.016}$ & $14.77^{+0.03}_{-0.03}$ & $13.45^{+0.03}_{-0.03}$ & $11.92^{+0.03}_{-0.03}$ & -- & -- & -- \\
CoKu Tau 4 & 58835.1969970 & $18.14^{+0.07}_{-0.07}$ & $16.525^{+0.015}_{-0.015}$ & $14.79^{+0.03}_{-0.03}$ & $13.48^{+0.03}_{-0.03}$ & $11.97^{+0.03}_{-0.03}$ & 0.074 & $0.073^{+0.024}_{-0.019}$ & $0.08^{+0.08}_{-0.05}$ \\
CoKu Tau 4 & 58838.2025996 & $18.07^{+0.07}_{-0.07}$ & $16.477^{+0.015}_{-0.015}$ & $14.72^{+0.03}_{-0.03}$ & $13.40^{+0.03}_{-0.03}$ & $11.96^{+0.03}_{-0.03}$ & 0.088 & $0.087^{+0.028}_{-0.021}$ & $0.09^{+0.08}_{-0.06}$ \\
CW Tau & 58819.4411022 & $12.99^{+0.07}_{-0.07}$ & $12.585^{+0.017}_{-0.017}$ & $11.43^{+0.04}_{-0.04}$ & $10.56^{+0.03}_{-0.03}$ & $9.80^{+0.04}_{-0.04}$ & 4.8 & $4.8^{+0.6}_{-0.5}$ & $6.4^{+2.5}_{-1.7}$ \\
CW Tau & 58824.1460621 & $13.10^{+0.07}_{-0.07}$ & $12.465^{+0.017}_{-0.017}$ & $11.24^{+0.04}_{-0.04}$ & $10.50^{+0.03}_{-0.03}$ & $9.75^{+0.04}_{-0.04}$ & 4.4 & $4.4^{+0.5}_{-0.5}$ & $6.0^{+2.2}_{-1.6}$ \\
CW Tau & 58827.1446019 & $12.42^{+0.07}_{-0.07}$ & $12.083^{+0.017}_{-0.017}$ & $10.94^{+0.04}_{-0.04}$ & $10.11^{+0.04}_{-0.04}$ & $9.31^{+0.05}_{-0.05}$ & 8.0 & $8.0^{+0.8}_{-0.8}$ & $10.7^{+4}_{-2.8}$ \\
... & ... & ... & ... & ... & ... & ... & ... & ... & ... \\
\enddata
\tabletypesize{\normalsize}
\tablecomments{Summary of our UBVRI photometry and $\dot{M}$ measurements obtained using the LDT during December 2019. Observing times are reported in MBJD in the TDB time frame. $\rm{\dot{M}_{fixed}}$ is calculated directly from our assumed stellar parameters. A discussion on photometric uncertainties is included in \S~\ref{ss:ldt_obs}. $\rm{\dot{M}_{rand}}$ and the associated uncertainties were found by re-sampling measurement uncertainties, while those for $\rm{\dot{M}_{sys}}$ include re-sampling both measurement uncertainties and system parameters. The reported values for $\rm{\dot{M}_{rand}}$ and $\rm{\dot{M}_{sys}}$ are the $50^{\rm{th}}$ percentiles, while the uncertainties correspond to the $16^{\rm{th}}$ and $84^{\rm{th}}$ percentiles. A complete machine readable table is available online. } 

\end{deluxetable*}
\begin{deluxetable*}{ccccc}
\centering 
\tablecolumns{5}
\tablecaption{LDT U-band Monitoring and $\dot{M}$ Measurements of GM Aur \label{tab:gmaur_Uband}}
\tablehead{\colhead{Exposure start} & \colhead{U} & \colhead{$\rm{\dot{M}_{fixed}}$} & \colhead{$\rm{\dot{M}_{rand}}$} & \colhead{$\rm{\dot{M}_{sys}}$} \\ \colhead{[MBJD]} & \colhead{[mag]} & \colhead{{\scriptsize$\mathrm{[10^{-8}M_\odot/yr]}$}} & \colhead{{\scriptsize$\mathrm{[10^{-8}M_\odot/yr]}$}} & \colhead{{\scriptsize$\mathrm{[10^{-8}M_\odot/yr]}$}} }
\startdata
58819.3318491 & $13.509^{+0.11}_{-0.11}$ & 0.9637 & $0.96^{+0.12}_{-0.10}$ & $1.0^{+0.5}_{-0.4}$ \\
58819.3322387 & $13.522^{+0.11}_{-0.11}$ & 0.9500 & $0.95^{+0.11}_{-0.11}$ & $1.0^{+0.5}_{-0.4}$ \\
58819.3326285 & $13.519^{+0.11}_{-0.11}$ & 0.9530 & $0.95^{+0.12}_{-0.10}$ & $1.0^{+0.5}_{-0.4}$ \\
58819.5011390 & $13.542^{+0.09}_{-0.09}$ & 0.9296 & $0.93^{+0.10}_{-0.09}$ & $0.9^{+0.5}_{-0.4}$ \\
58819.5017626 & $13.538^{+0.10}_{-0.10}$ & 0.9337 & $0.94^{+0.11}_{-0.10}$ & $1.0^{+0.5}_{-0.4}$ \\
58819.5023838 & $13.550^{+0.18}_{-0.18}$ & 0.9220 & $0.91^{+0.19}_{-0.16}$ & $0.9^{+0.6}_{-0.4}$ \\
58819.5030051 & $13.532^{+0.11}_{-0.11}$ & 0.9398 & $0.94^{+0.12}_{-0.10}$ & $1.0^{+0.5}_{-0.4}$ \\
58819.5036263 & $13.570^{+0.27}_{-0.27}$ & 0.9023 & $0.93^{+0.3}_{-0.26}$ & $0.9^{+0.6}_{-0.4}$ \\
... & ... & ... & ... & ...
\enddata
\tablecomments{The scatter in adjacent values of $\rm{\dot{M}}$, which is typically on the order of $0.01$ to $0.03 \times10^{-8} \, \mathrm{M_\odot \, yr^{-1}}$, is a better estimate of true random uncertainty since it is set by the internal precision of our differential photometry. Note that some of that scatter may be caused by real variability. A complete machine readable table is available online. }
\end{deluxetable*}

\subsection{\textit{HST} \label{ss:HST_obs}}

\begin{figure}[t]
    \centering
    \includegraphics[width = .99\linewidth]{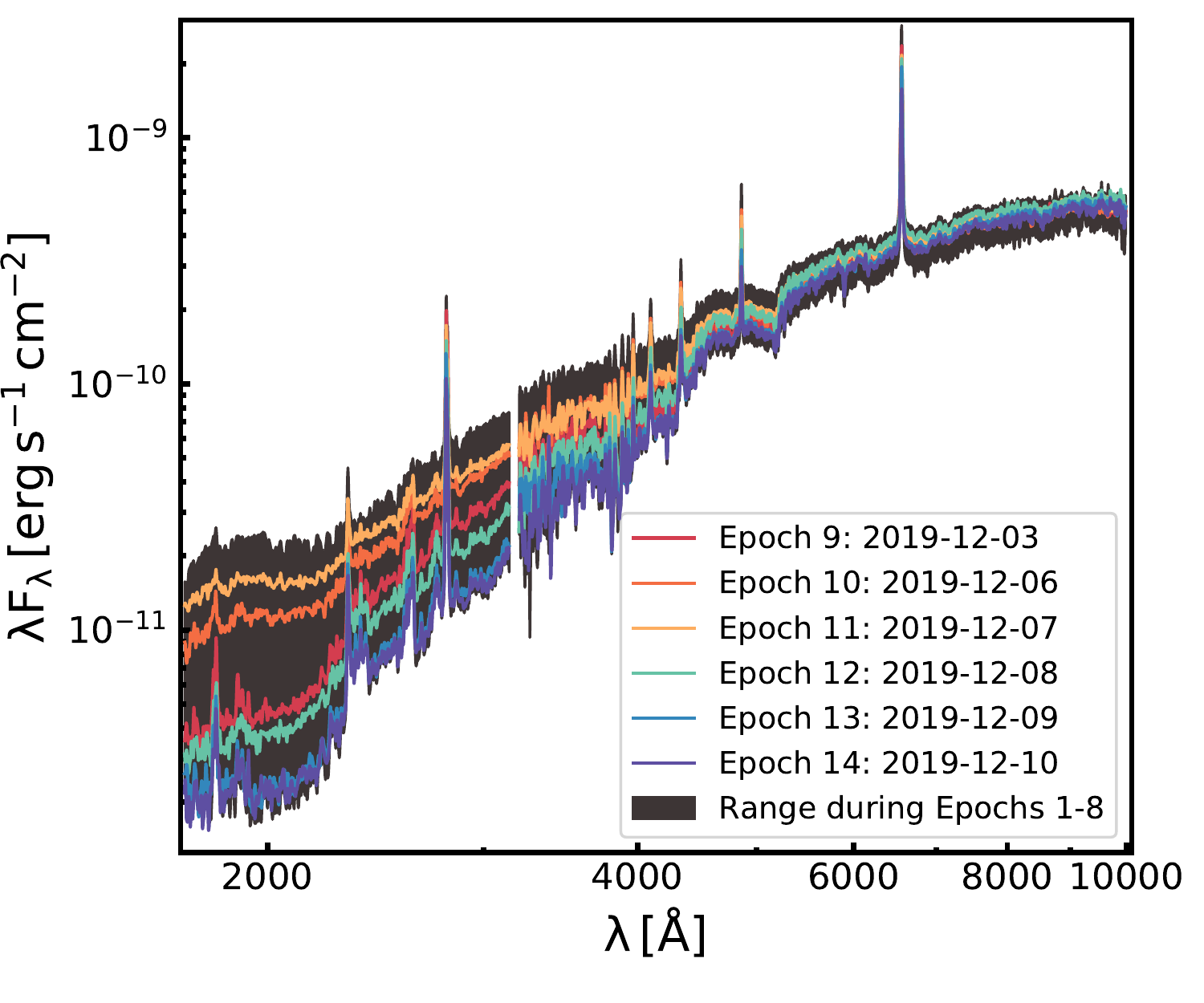}
    \caption{Six epochs (Epochs 9 - 14) of \HST STIS spectra with contemporaneous NUV - NIR coverage of GM Aur observed during December 2019. The dark solid region represents the range between the maximum and minimum values observed during the previous \HST STIS observations (Epochs 1 - 8) of GM Aur presented in \citet{ingleby15} and \citet{robinson19}. For presentation purposes only, we show the spectra smoothed with a Gaussian Kernel.}
    \label{fig:gmaur_spectra}
\end{figure}

We present six NUV - NIR ($1700 - 10000$ \angstrom) low resolution ($R \approx 500 - 1000$) spectra of GM Aur using STIS aboard \textit{HST}, with a 1-3 d separation between epochs. Analysis for five out of six of these observations was first presented in \citet{espaillat21}. Here we focus on making connections between these spectra and the simultaneous \TESS and contemporaneous LDT observations, as well as add a new epoch and an additional sub-exposure analyses. We refer to these observations as Epochs 9 - 14 to remain consistent with the eight previous observations presented in \citet{ingleby15} and \citet{robinson19}. Epoch 9 (2019-12-03) is the newly added epoch. This observing strategy was designed to cover a full rotation period of GM Aur \citep[6.1 d;][]{percy10} and to be simultaneous with \textit{TESS}. The \HST visits were originally scheduled to be daily, but had to be rescheduled due to an \HST instrument error causing the gap in observations between Epochs 9 and 10.  Epochs 9, 10, 11, 12, and 13 were taken simultaneously with the \TESS observations. Epoch 14 (2019-12-10) fell into the data downlink gap during the perigee, and thus are contemporaneous rather than simultaneous with the \TESS dataset.  None of our \HST observations occurred perfectly simultaneously with our LDT observations. Table \ref{tab:hst_summary} summarizes the six \HST STIS observations of GM Aur. Figure \ref{fig:gmaur_spectra} shows our six epochs of \HST STIS spectra overlaid on the range of the eight previous \HST STIS observations. These observations were obtained during the \HST Director's Discretionary time GO program 16010 (PI: C. Robinson). 

We took the spectra using the $52"\times2"$ slit and the G230L grating with the NUV-MAMA detector and the G430L and G750L gratings with the CCD detector. All three spectral orders were taken sequentially within a single orbit, making the observations contemporaneous to within roughly 50 minutes. Moderate fringing occurs at the red end of the NIR spectrum. To correct this, we follow the standard procedure described in \citet{goudfrooij98} using a contemporaneous fringe flat taken alongside the observations during terrestrial occultation. 

The NUV data were taken in the time-tag mode of the NUV-MAMA detector, which records individual photon arrival times, making it possible to break a single long exposure into several shorter sub-exposures. While most of the analysis in this work relies on the full exposure length for the NUV observations, we take advantage of this in \S\ref{ss:gmaur_timetag} to study minute-timescale changes in the accretion behavior. The initial delay and exposure times for these sub-exposures were chosen such that they are exactly simultaneous with overlapping \TESS exposures (see \S\ref{ss:time} for details on matching time frames between \textit{HST}, \textit{TESS}, and LDT). 

To isolate emission arising from accretion, we require an accurate model of the emission arising from the non-disturbed stellar photosphere. Non-accreting WTTS are preferred as spectral templates due to significant chromospheric emission in young stars in the UV \citep[see][]{ingleby11}. A NUV - optical \HST STIS spectrum of the K5 \citep{luhman04} WTTS RECX 1 was used as a spectral template in the analysis of GM Aur. This spectrum was taken with the matching 52" x 2" slit and the G230L and G430L gratings using the CCD detector during the \HST GO program 11616 (PI: G. Herczeg).  

\begin{deluxetable*}{cccccc}
\centering
\tablecolumns{5}
\tablecaption{HST Observation Summary \label{tab:hst_summary}}
\tablehead{Target & Epoch & Proposal ID & Gratings & Date [UT] & G230L Exposure start }
\startdata
GM Aur & 9 & 16010 & G230L, G430L, G750L & 2019-12-03 &  58820.4834602  \\
GM Aur & 10 & 16010 & G230L, G430L, G750L & 2019-12-06 & 58823.4655720  \\
GM Aur & 11 & 16010 & G230L, G430L, G750L & 2019-12-07 & 58824.4593470 \\
GM Aur & 12 & 16010 & G230L, G430L, G750L & 2019-12-08 & 58825.4350766 \\
GM Aur & 13 & 16010 & G230L, G430L, G750L & 2019-12-09 & 58826.4345312 \\
GM Aur & 14 & 16010 & G230L, G430L, G750L & 2019-12-10 & 58827.6387985 \\
RECX 1 & -- & 11616 & G230L, G430L        & 2010-01-22 & 55218.4034028${}^\dagger$  \\
\enddata
\tablecomments{Summary of the \HST  STIS observations of the CTTS GM Aur and the WTTS RECX 1. Each visit was completed within a single orbit. The epoch naming scheme is adopted from \citet{ingleby15} and \citet{robinson19}, which discuss epochs 1-3 and epochs 4-8 in more detail, respectively.  Epochs 10-14 were first presented in \citet{espaillat21}. ${}^\dagger$While the other exposure start times are in MBJD in the TDB timescale, the exposure time start for RECX 1 is listed in MJD in UTC because simultaneity is not required.}
\end{deluxetable*}

\subsection{Transforming to a uniform time scale \label{ss:time}}
Because this is a multi-observatory study with importance placed on simultaneity, the timestamps for each observation must be placed into the same time scale. We choose to adopt the Barycentric Dynamical Time (TDB) scale and account for photon travel time to the barycenter of the solar system and present dates as Modified Barycentric Julian Dates. For \textit{TESS}, timestamps were transformed into this system using the \texttt{Lightkurve} package \citep{lightkurve18}. For \textit{HST}, this was accomplished by following the steps laid out in \citet{dashevsky00} using the orbital information of \HST during epochs 9-14. Finally, the LDT data were transformed to this time scale using the \texttt{astropy.time} package \citep{astropy18}.

\section{Analysis \label{s:analysis}}

\subsection{Stellar parameters \label{ss:stellar_params}}
With the exception of GM Aur, we adopt stellar parameters for all of our objects from \citet{herczeg14}. Those authors include a veiling continuum emission from accretion and extinction during their spectral classification process, which are both critical for accurate measurements of the accretion rate. Spectral types were converted into effective stellar temperatures by interpolating between the grid of WTTS presented in Tab. 4 of \citet{herczeg14}. Following that work, we include K8 as an intermediate spectral type between K7 and M0. We scaled the stellar luminosities from the distances that were assumed in that work to those derived from parallaxes from \Gaia EDR3 \citep{gaia16, gaia21a} and obtained new stellar radii based on this correction. Stellar masses were obtained by interpolating between the Mesa Isochrones and Stellar Tracks (MIST) model grid \citep{choi16, dotter16}. The updated stellar parameters for each object are presented in Table \ref{tab:stellar_params}.

As discussed by \citet{garufi19}, the parallax from \Gaia DR 2 for RY Tau is known to be inconsistent with previous measurements and is assumed to be erroneous. However, this appears to have been corrected with the release of \Gaia EDR3 with a value of 138.2 pc, which we adopt. This new distance is roughly consistent with the previous \textit{Hipparcos} measurement of 133 pc \citep{esa97} which has been adopted by previous works. For GM Aur, we adopt the stellar parameters from \citet{manara14} to remain consistent with the analysis of the \HST STIS spectra presented in \citet{robinson19, espaillat19a, espaillat19b, espaillat21, espaillat22}. 

\begin{deluxetable*}{cccccccccccc}
\centering 
\tablecolumns{3}
\tablecaption{Adopted Stellar Parameters \label{tab:stellar_params}}
\tablehead{Object & 2MASS  & SpT & $\rm{A_v}$ & $\rm{T_\star}$ & $\rm{L_\star}$ & Distance & $\rm{R_\star}$ & $\rm{M_\star}$ & \it{i} & Disk Morpology & Refs. \\ 
& & & $[\rm{mag}]$ & $[\rm{K}]$ & $[\rm{L_\odot}]$ & $[\rm{pc}] $ & $[\rm{R_\odot}]$ & $[\rm{M_\odot}]$ & $[{}^\circ]$ & & }
\startdata
BP Tau & J04191583+2906269 & M0.5 & 0.45 & 3900 & 0.39 & 127.4 & 1.36 & 0.66 & $38.2^{+0.5}_{-0.5}$ & F & [a]\\
CoKu Tau 4 & J04411681+2840000 & M1.1 & 1.75 & 3720 & 0.39 & 155.1 & 1.52 & 0.49 & $30^{+20}_{-19}$ & T & [b,c]\\
CW Tau & J04141700+2810578 & K3 & 1.8 & 4543 & 0.45 & 131.5 & 1.09 & 0.97 & $65^{+2}_{-2}$ & F & [d, e] \\
CY Tau & J04173372+2820468 & M2.3 & 0.35 & 3560 & 0.25 & 126.3 & 1.30 & 0.4 & $32^{+1}_{-1}$ & F & [f, m]\\
DD Tau & J04183112+2816290 & M4.8 & 0.75 & 3190 & 0.27 & 126.7 & 1.71 & 0.19 & -- & F & [g, h]\\
DE Tau & J04215563+2755060 & M2.3 & 0.35 & 3560 & 0.50 & 128.0 & 1.87 & 0.38 & $66^{+7}_{-7}$ & F & [e, f] \\
DS Tau & J04474859+2925112 & M0.4 & 0.25 & 3900 & 0.24 & 158.4 & 1.07 & 0.72 & $65.2^{+0.3}_{-0.3}$ & PT & [a, i] \\
FM Tau & J04141358+2812492 & M4.5 & 0.35 & 3190 & 0.07 & 132.0 & 0.88 & 0.18 & $55^{+2}_{-2}$ & F & [g,h]\\
FN Tau & J04141458+2827580 & M3.5 & 1.15 & 3410 & 0.52 & 129.9 & 2.06 & 0.29 & $20^{+10}_{-10}$ & F & [f,j]\\
FO Tau & J04144928+2812305 & M3.9 & 2.05 & 3410 & 0.57 & 136.1 & 2.17 & 0.29 & $39^{+42}_{-42}$ & F & [h]\\
GM Aur & J04551098+3021595  & K5 & 0.6 & 4350 & 1.28 & 159.6 & 2.0 & 1.36 & $56.45^{+0.06}_{-0.05}$ & T & [k]\\
RY Tau & J04215740+2826355 & G0 & 1.85 & 5930 & 11.9 & 138.2 & 3.28 & 2.13 & $65^{+0.1}_{-0.1}$& T & [a, i]\\
UY Aur & J04514737+3047134 & K7.0 & 1.0 & 4020 & 1.01 & 152.3 & 2.07 & 0.63 & $23.5^{+7.8}_{-6.6}$ & F & [a] \\
V819 Tau & J04192625+2826142 & K8.0 & 1.0 & 3960 & 0.47 & 129.3 & 1.45 & 0.70 & $46^{+30}_{-30}$ & D & [c,l] \\
\enddata
\tablecomments{Stellar parameters are from \citet{herczeg14} and \citet{gaia21a}, while masses are derived using the MIST stellar evolution models. For GM Aur, we choose to adopt values from \citet{manara14} to be consistent with \citet{robinson19}, and \citet{espaillat21}. The disk morphologies of F, T, PT, and D refer to full, transitional, pre-transitional, and debris, respectively. These are derived primarily from mm observations, but we use IR SED literature classification/modeling when resolved mm observations are not available (DD Tau and FM Tau). We are not aware of any literature morphology classifications for FN Tau, so we estimate it here by-eye based on the 228 GHz continuum images presented in [f]. References for inclination and disk morphology: [a]~\citet{long19}, [b]~\citet{ireland08}, [c]~\citet{ballering19}, [d]~\citet{bacciotti18}, [e]~\citet{pietu14}, [f]~\citet{simon17}, [g]~\citet{furlan11}, [h]~\citet{akeson19}, [i]~\citet{long18a}, [j]~\citet{kudo08}, [k]~\citet{macias18}, [l]~\citet{hardy15}, and [m]~\citet{perez15}.}
\end{deluxetable*} 

\subsection{Measuring \texorpdfstring{$\dot{M}$}{mdot} from U-band excess \label{ss:uband_mdot}}
Excess emission above photospheric levels from accretion in the wavelengths covered by the broadband U filter can be used to measure $\rm{\dot{M}}$ through empirical relationships \citep[e.g.,][]{gullbring98}. For each object (except the LDT GM Aur monitoring efforts and \HST observations, see Section~\ref{ss:HST_obs}), we approximate the photospheric contribution to the observed U-band flux for each night and object by scaling BT Settl photospheric models \citet{allard14} using the stellar parameters (see Section \ref{ss:stellar_params}) and convolving the resulting spectrum with the U transmission curve.  After subtracting the photospheric emission, the excess U-band accretion luminosity is converted to a bolometric accretion luminosity using the empirical relationship from \citet{robinson19}. This is then converted into a mass accretion rate by assuming that the material is falling at free-fall velocities from an inner magnetospheric radius of $5 \rm{R_\star}$. This value of  $5 \rm{R_\star}$ arises from balancing the ram pressure from the Keplerian motion of the disk and the magnetic pressure for typical CTTS parameters \citep[see][]{hartmann16}. Recent interferometric observations of the magnetosphere of TW Hya found similar results, with a measured of radius of $3.5 \rm{R_\star}$ \citep{gravity20}. Additionally, because of the weak scaling between accretion luminosity and magnetospheric truncation radius ($ L_{acc} \propto \big(1 - \frac{R_\star}{R_{M}}\big)$) due to the flow approaching the free-fall velocity, small differences  in $R_{M}$ between objects will not strongly influence our results. 

To estimate uncertainties on our reported mass accretion rates, we use a Monte Carlo approach. We assume that the uncertainty distributions for our parameters can be approximated as Gaussian probability distributions. Following \citet{herczeg14}, we adopt uncertainties of $\rm{\sigma_{A_v} = 0.2}$ (appropriate for moderately veiled stars) and a $0.1$ dex uncertainty for $\rm{L_\star}$. For effective temperature, we adopt an uncertainty of $\rm{\sigma_{T_{eff}}} = 100$ which is approximately 1 spectral class. When sampling $\rm{L_\star}$ and $\rm{T_\star}$, we apply a Gaussian prior on the age of the cluster of 2.0 Myr \citep{kenyon95} with a width of 1 Myr. The inclusion of this prior shifts the center of the re-sampled distributions of $\rm{L_\star}$ and $\rm{T_\star}$ and thus the distribution of $\rm{\dot{M}}$. We found that this effect was particularly noticeable for stars that \citet{herczeg14} identified as being heavily veiled (e.g., CW Tau, DS Tau, FM Tau). While we have assumed that a single age is representative of the age of the Taurus star-forming region \citep[e.g.,][]{kenyon95, luhman18}, other authors suggest that it may contain older populations of stars \citep{kraus17, krolikowski21}. With this in mind, we explored an alternative age prior of the functional form of $\frac{1}{2} \Big(1 -\rm{erf}(\frac{A - A_0}{\sqrt2 \omega})\Big)$, which results in roughly equal probability before encountering a drop-off with a characteristic width $\rm{\omega}$ centered at $\rm{A_0}$. With $\rm{A_0 = 15}$ Myr and $\omega = 2$ Myr, we find good agreement between the mean of our sampled distribution and the the stellar parameters from \citet{herczeg14}. However, because the shift in sampled stellar parameters primarily occurs for objects that are heavily veiled \citep[which should be an indicator of  youth][]{hartmann98}, we suggest that the shifts in the $\rm{L_\star}$ and $\rm{T_\star}$ distributions that appear under the Gaussian prior should not be used as evidence for an older population, and are instead simply an indication of the difficulty in accurately measuring stellar parameters for strongly accreting sources. We thus chose to adopt the single-age Gaussian prior and also note that the ultimately this choice of prior does not strongly influence our interpretation of our results.

Additionally, we re-sample our U-band photometry, \textit{GAIA} parallaxes, and the coefficients in the empirical linear relationship between $\rm{L_U}$ and $\rm{L_{acc}}$ from \citet{robinson19} using their associated uncertainties. Values of $\rm{M_\star}$ for each re-sampling of $\rm{L_\star}$ and $\rm{T_\star}$ are found by interpolating within the MIST model grid \citet{choi16, dotter16}. When sampling stellar masses lower than 0.09 $\rm{M_\star}$, we instead interpolate within the \citet[][hereafter BHAC]{baraffe15} evolutionary tracks. From this analysis, we found that typical uncertainties are on the order of 0.2 dex of the reported $\rm{\dot{M}}$ values. We refer to the $50^{th}$ percentile of this distribution as $\rm{\dot{M}_{sys}}$. We do not report values of $\rm{\dot{M}_{sys}}$ for RY Tau because both its previously measured $\rm{T_\star}$, $\rm{L_\star}$ values lie near the upper edge of the MIST model grid, and would require either extrapolation or rejection of points when sampling, each of which may introduce biases.

If we exclude uncertainties on $\rm{A_v}$, $\rm{T_{eff}}$, $\rm{D}$, $\rm{L_\star}$ (i.e., systematic effects that would be eliminated by studying an individual star), then typical uncertainties are instead on the order of 0.09 dex. We refer to these reported values as $\rm{\dot{M}_{rand}}$. Note that our uncertainty estimates for this value are still significantly larger than what the scatter in our U-band photometry alone would suggest because we include the uncertainty on the relationship between $\rm{L_{acc}}$ and $\rm{L_U}$. Finally, we calculate the mass accretion rate directly from our stellar parameters and data without resampling, which we refer to as $\rm{\dot{M}_{fixed}}$. Each are reported in Table~\ref{tab:LDT_table}. In most cases, these values agree with each other. The largest exception is RY Tau in low accretion states, where it becomes difficult to measure the accretion excess against its bright continuum. We exclude negative accretion rates as being nonphysical, which shifts the $50^{\rm{th}}$ percentile to higher values during low accretion states.

\citet{herczeg14} could not resolve the four known binaries in our sample (UY Aur, Coku Tau 4, DD Tau, and FO Tau), and thus their stellar parameters arise from their combined spectra. While this should not introduce large systematic errors into our $\rm{L_{acc}}$ measurements, it may affect our resulting measurement of $\rm{\dot{M}}$ if $\rm{M_\star/R_\star}$ varies significantly from that of our single star approximation. Under the assumption that accretion is occurring primarily onto the primary star, the most extreme deviation from our single star approximation will occur for two equally sized stars. Under the MIST/BHAC models, this results in a systematic multiplicative underestimation of the mass accretion rate for these objects of up to about $30\%$. We do not include this effect in the reported systematic uncertainties for $\rm{\dot{M}_{sys}}$. This effect does not significantly change the interpretation of our results.

\subsection{Measuring \texorpdfstring{$\dot{M}$}{mdot} from \textit{HST}}
We take a different approach for GM Aur because of our additional simultaneous \textit{HST} observations and ground-based U-band monitoring. To probe the accretion columns of GM Aur, the mass accretion rate and surface coverage of accretion shocks with different densities were measured with the \HST STIS spectra using the accretion shock models and fitting methods of \citet{robinson19}. These models are an updated version of the models of \citet{calvet98} and the modeling efforts for five out of six of these epochs of GM Aur were first presented in \citet{espaillat21}. The model and fitting methods are briefly described here.

The structure and emission arising from the post-shock region is solved under a given kinetic energy flux, $F_i = \frac{1}{2} \rho u^3$ where $\rho$ is density and $u$ is velocity. These models work under the assumption that the accretion flow is in freefall in the pre-shock region, making the kinetic energy flux directly proportional to the density of the column. Half of the emission is radiated toward the star, where it heats the underlying photosphere, and the other half irradiates the pre-shock region. The outgoing, reprocessed emission from these two regions is then summed and scaled by a multiplicative filling factor, $f_i$, which is treated as a free parameter. This is repeated for each value of $F_i$. A non-accreting WTTS, multiplied by a free parameter scaling factor, $s$, is used as a template for the non-accreting regions of the stellar photosphere. The scaling factor $s$ is defined as the ratio between the brightest observation and the photospheric template as measured through the V filter. The emission from each component is then summed to produce an estimate of the continuum emission. 

\begin{deluxetable}{ccccc}[b]
\centering 
\tablecolumns{5}
\tablecaption{Light Curve Morphology and Rotation Periods \label{tab:QMT}}
\tablehead{Object & Q & M & Period & Variability \\ & & & $[d]$ & Classification}
\startdata
BP Tau & 0.83 & -0.03 & $4.39^\dag$ & QPS \\
CoKu Tau 4 & 0.33 & -0.14 & 1.91/3.02 & MP (QPS) \\
CW Tau & 0.56 & -0.10 & 9.40 & QPS \\
CY Tau & 0.38 & 0.50 & 3.99 & QPD \\
DD Tau & 0.85 & -0.49 & $5.95^\dag$ & B \\
DE Tau & 0.39 & -0.06 & 5.79 & QPS \\
DS Tau & 0.36 & -0.27 & 8.65 & B \\
FM Tau & 0.70 & -0.02 & $5.37^\dag$ & QPS \\
FN Tau & $< 0.6$ & 0.03 & 8.65 & P/QPS \\
FO Tau & 0.32 & -0.29 & 4.18 & B \\
GM Aur & 0.46 & 0.02 & 5.79 & QPS \\
RY Tau & 0.82 & 0.61 & $3.33^\dag$ & QPD \\
UY Aur & 0.74 & -0.49 & 3.04 & B \\
V819 Tau & 0.03 & -0.27 & 5.48 & P (QPS)
\enddata
\tablecomments{Light curve periodicity ($Q$), symmetry ($M$) \citep[see][]{cody14}, inferred rotation period, and light curve morphology classification as measured from our \TESS data. The classifications in parentheses for CoKu Tau 4 and V819 Tau are those determined from the Q and M metrics which do not match our by-eye classification. The light curve of FN Tau is quite noisy, resulting in an upper limit on Q and us assigning a by-eye classification of either P or QPS. ${}^\dag$Inferred periods for objects that display strong aperiodic behavior may not be accurate.} 
\end{deluxetable}

In this approach, the photosphere is assumed to remain constant between epochs, i.e., the value of $s$ is multiplied uniformly by the photospheric template for each epoch. Treating $s$ as a single free parameter requires that $f_i$ for all epochs and $s$ be fit simultaneously. To identify models that fit the observations well in this high-dimensional parameter space, posteriors for $f_i$, $s$, and model uncertainty $w$ were sampled using a Markov Chain Monte Carlo approach. In particular, we adopt the affine-invariant \citet{goodman10} sampling algorithm via the Python-based ensemble sampler \texttt{emcee} \citep{foreman-mackey13}. The model uncertainty, $w$, is treated as a nuisance parameter and can be marginalized. Each parameter is evaluated in log-space (limiting possible values to positive quantities) and we impose priors such that $s$, $w$, and $f_i$ cannot be larger than 1 to remain physical.  Additionally, we impose a Gaussian prior on $s$ with a width of 0.1 centered on the value of $s$ derived from the analysis of the Epochs 1-8 from \citet{robinson19}.  It is important to note that the resulting reported posteriors do not include the systematic uncertainty arising from the stellar parameters and extinction. \citet{robinson19} found an additional 0.04 dex systematic uncertainty in the fitted parameters is appropriate in most cases. For more discussion on the fitting technique and the model, see that work, and \citet{calvet98}. We also adopt the scaled template when calculating mass accretion rates using our U-band monitoring data for GM Aur. When resampling the photosphere in our Monte Carlo simulation, we assume an uncertainty of $10\%$ based on the results of \citet{robinson19}.

\subsection{\TESS Symmetry and periodicity \label{ss:QM}}
Q and M are statistical metrics that measure periodicity and symmetry around the mean of a light curve, respectively. We measured these metrics for each for object within our sample using their \TESS light curves. The Q and M metrics were developed by \cite{cody14} and have been used to characterize several star-forming regions \citep[e.g., NGC 2264, Ophiuchus,][]{cody14, cody18}. A recent study also presented Q and M values measured from K2 light curves of the Taurus star-forming region \citep{cody22}. We note that there is no overlap between our sample and objects presented in that work.
These metrics can be used to separate objects into empirical variability classes.
The classifications include burster (B), purely periodic (P), quasi-periodic symmetric (QPS), stochastic (S), quasi-periodic dipper (QPD) and aperiodic dipper (APD). In addition to these classifications, we also note the possibility of several other variability classes, including multi-periodic (MP), which would not be identified by these metrics. We adopt the regions of Q and M parameter space as defined by \citet{cody18} to determine variability classifications for the objects. More specifics on the algorithms used to measure these metrics can be found in that work and \citet{cody14}. 

\subsection{Measuring periods from \TESS}
One of the steps for measuring Q requires determining the stellar rotation period. Following \citet{cody14}, we measured the period by interpolating the \TESS data onto a uniform grid and autocorrelating the data. The first significant peak of the autocorrelation function is then identified. Next, we applied a Lomb-Scargle periodogram \citep{lomb76, scargle82}. The resulting periodogram was then weighted by a Gaussian centered at the peak of the autocorrelation function with a width of $30\%$ of the peak period. This was found to more accurately recover rotational periods of models of accreting young stars \citep{robinson21} when compared to the box-car scheme of \citet{cody14}. We took a different approach for CoKu Tau 4, since by-eye inspection of its light curve reveals that it resembles a MP binary. Instead, we identified the two strongest periods directly using a Lomb Scargle periodogram without any additional weighting. We note that our reported periods for sources that exhibit very strong aperiodic variability may not be reliable (BP Tau, DD Tau, RY Tau, and FM Tau).

\begin{figure}[h]
    \centering
    \includegraphics[width = \linewidth]{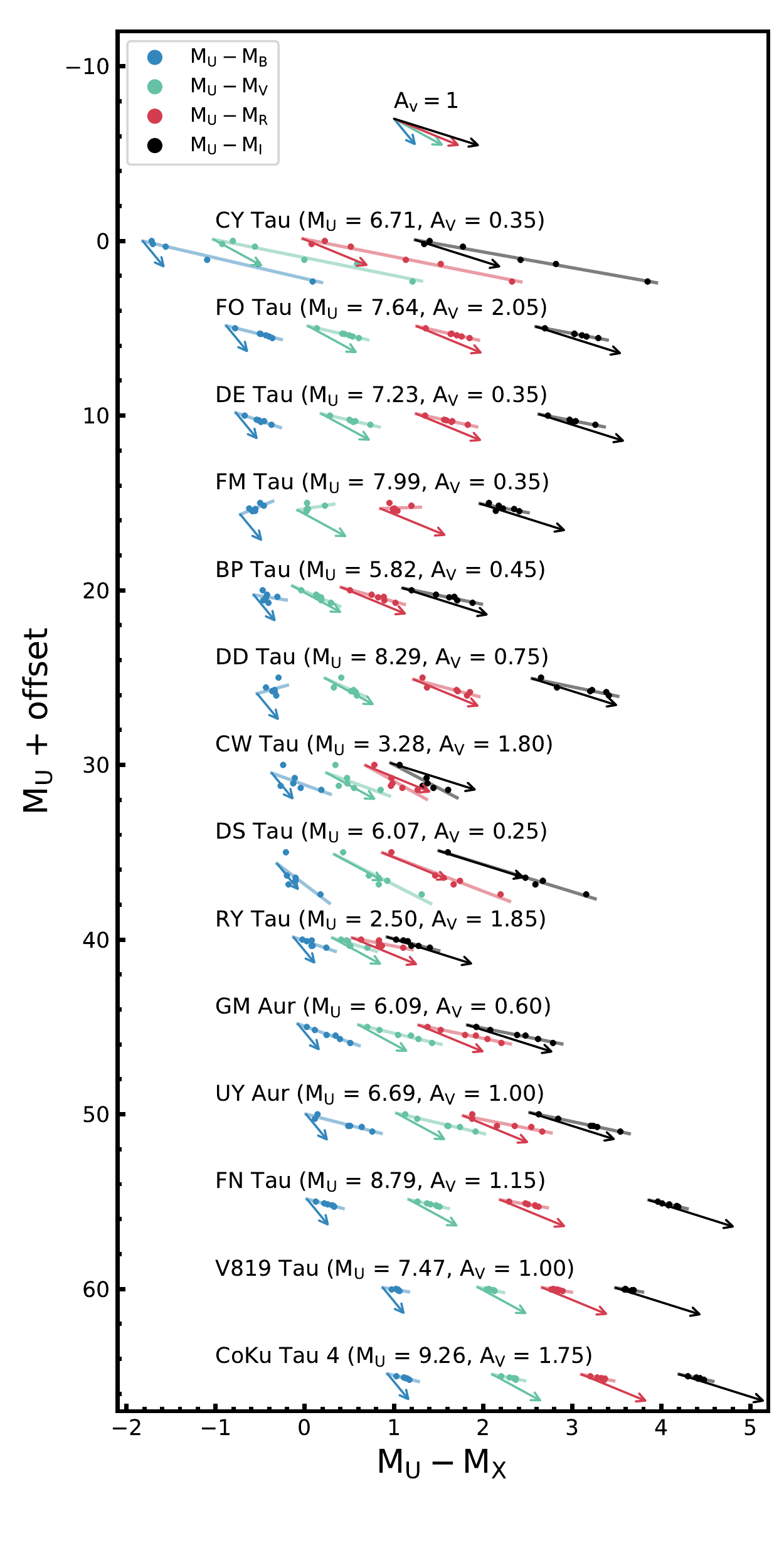}
    \caption{$\rm{M_U-M_B}$, $\rm{M_U-M_V}$, $\rm{M_U-M_R}$, and $\rm{M_U-M_I}$ color-magnitude diagram of the 14 targets in the sample (sorted by the bluest observed $M_U - M_B$ color) from our ground-based LDT monitoring campaign. $\rm{M_U}$ is shown with an offset for clarity. A set of extinction vectors for $\rm{A_V} = 1$ is included and a linear fit to the data is present for comparison. The reported $\rm{M_U}$ values are the peak brightness for each object. }
    \label{fig:cmd}
\end{figure}

\section{Results \label{s:results}}

\subsection{LDT color magnitude diagrams}
We correct for interstellar reddening in our LDT images using the extinction law of \citet{cardelli89} ($\rm{R_V} = 3.1$). Using the distances from \textit{GAIA} we convert our apparent magnitudes into absolute magnitudes. The extinction-corrected, color-magnitude diagrams for each object are shown in Figure~\ref{fig:cmd}. The objects are shown sorted by their bluest $\rm{M_U - M_B}$ slope with an offset for clarity.
In Figure~\ref{fig:cmd}, we also include $\rm{A_v} = 1$ extinction vectors, which show the expected slope for changes induced by variability in extinction along the line of sight. If the observed variability was caused solely by extinction, we would expect our photometry to lie along those vectors.
We find that for most of the objects included in our sample, the $\rm{M_U - M_B}$, $\rm{M_U}$ slope is inconsistent with changes purely in the extinction along the line of sight. This is expected, since T Tauri stars are known to have variable accretion rates, the effect of which is most apparent at shorter wavelengths.

\subsection{Connecting \texorpdfstring{$\dot{M}$}{mdot} from U-band excess to \TESS}
\begin{figure*}
    \includegraphics[width = \linewidth]{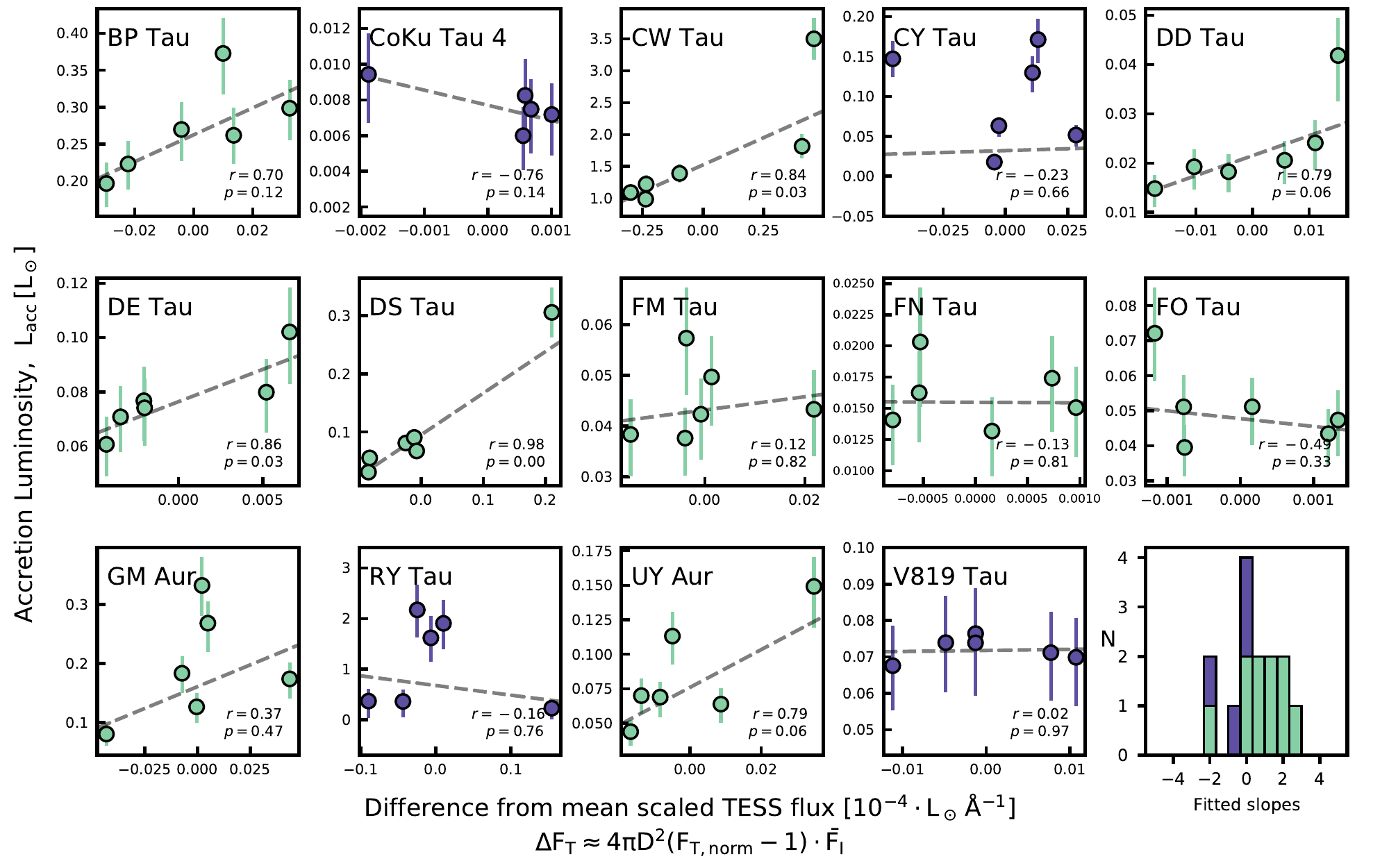}
    \caption{\label{fig:TESS-vs-Lacc} $\rm{L_{acc}}$ plotted against the changes ($\rm{\Delta F_T}$) in the normalized \TESS flux ($\rm{F_T,norm}$) scaled using our mean I-band photometry ($\bar{F}_I$) and the distance to the object ($D$). The Pearson correlation coefficient, $r$, and the associated $p$-value are included in the panel. The dashed line is a linear regression to show the trend. Green points represent sources with light curve variability classifications that are often associated with accretion (in our sample, bursters and quasi-periodic symmetric sources), while purple points represent objects with those thought to be primarily driven by non-accretion processes (in our sample, dippers, multiperiodic, and periodic sources). \textit{Bottom right}: Histogram of the slopes from each source. This figure excludes the continuous U-band monitoring of GM Aur. (see Figure~\ref{fig:GMAur-TESS-vs-mdot}).}
\end{figure*}

We leveraged our multi-wavelength coverage to search correlations between the \TESS light curves and accretion. To do this, we identify the \TESS exposures that are simultaneous with our U and I photometry.  We then deredden our I-band photometry, and convert from magnitudes to flux ($F_{I,\lambda}$) by adopting standard zeropoints. We then make the rough approximation of scaling the mean of the simultaneous \TESS exposures to the mean of the dereddened, I-band fluxes. We chose I band because both it and the \TESS bandpass are centered near 8000 $\angstrom$. We then subtract off the mean scaled \TESS flux and correct for the distance and compare the result to the accretion luminosity measured through the U-band excess in Figure~\ref{fig:TESS-vs-Lacc}. 

Within Figure~\ref{fig:TESS-vs-Lacc}, we separate objects with variability classes (see Table~\ref{tab:QMT}) that are associated with accretion variability (in our sample QPS, B) from those with variability arising primarily from other sources such as occultation and spot modulation (in our sample, QPD, MP, P). However, accretion and these other drivers of variability can and do occur simultaneously, which results in chimeric light curves with features arising from multiple mechanisms. This is particularly relevant for QPS objects, for which previous work has suggested that multiple sub-categories may exist even within the broader QPS category \citep{cody18} that are a mix of accretion and non-accretion processes. This introduces some amount of uncertainty in the source of the variability in our \TESS light curves and its connection to the inferred mass accretion rate. We do note that none of the regions flagged by-eye as stellar flares by their characteristic rapid rise and exponential decay in the \TESS light curves overlapped with our LDT observations, so we can eliminate that specific source of astrophysical noise as a significant contaminant. 

To test whether our inferred values of $\dot{M}$ and the variability in \TESS are related, we calculated Pearson correlation coefficients and their associated p-values and did a linear regression for each source. We find positive slopes for our fits for 10 out of 14 of the sources (but note that some of the objects display large p-values). For the objects that have negative slopes, only two have variability classifications that are typically associated with accretion variability (FN Tau, FO Tau). Both of these objects only have relatively small variations in the \TESS light curve compared to their mean brightness ($\sim 1\%$ and $5\%$, respectively) and have large p-values. For the objects that do have positive trends, we find significant scatter amongst the fitted slopes. For cases within our sample that have variability driven by accretion and other sources (particularly some QPS sources), we would expect that both the observed \TESS flux and the inferred mass accretion rate would be affected in part by some of the non-accretion related events. This could introduce a non-causal correlation between the two quantities. Previous work also identified differences in color slopes between stochastic accretors and dippers \citep{venuti15}, which would explain some of the scatter in the measured slopes between our inferred accretion rates and measured \TESS fluxes. Thus, we suggest that while \TESS is indeed capable of tracing some facets of the accretion process, it can be strongly influenced by other astrophysical sources noise. We discuss this further in Section \ref{ss:tess_and_mdot}. 

One caveat to this analysis is that the \TESS bandpass is significantly wider than the I bandpass, so differences in the spectral energy distribution between sources and the shape of the transmission curve may introduce systematic offsets. This does not affect the strength of the correlation coefficients for individual objects, but it may make comparing multiple objects more difficult. However, given the wide dispersion in fitted slopes, this effect is not large enough to strongly influence our interpretation. We explored this effect in part by repeating our analysis but instead normalize using our R-band photometry. Similar to I-band, we find large degrees of scatter in the measured slopes between $\rm{L_{acc}}$ and the scaled \TESS flux.

\subsection{GM Aur \label{sss:gmaur_ldt}}
\begin{figure*}
    \centering
    \includegraphics[width = 0.9\linewidth]{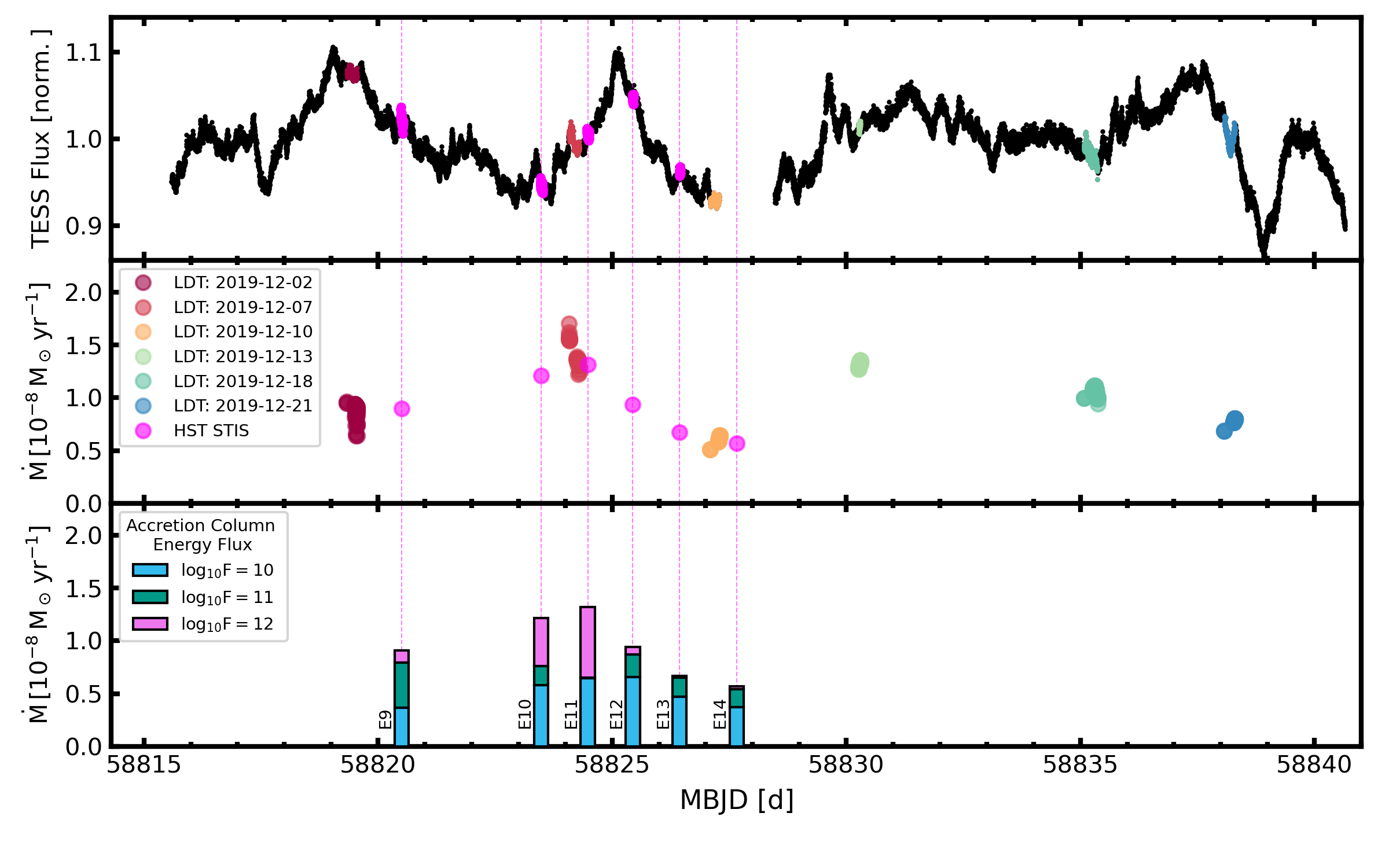}
    \caption{ {\textit{Top:}} Normalized \TESS light curve of GM Aur. Points with overlapping {\textit{HST}}/LDT measurements of $\rm{\dot{M}}$ are highlighted in color. The dashed magenta lines mark the middle of the \HST observations across all panels. \textit{Middle:} $\rm{\dot{M}}$ from the LDT U-band photometry and HST plotted as a function of time. Note that the peak of the measured $\rm{\dot{M}}$ is offset from the peak of the \TESS light curve.  \textit{Bottom:} The contribution to $\rm{\dot{M}}$ from the three accretion shock model components with different energy fluxes for each \HST epoch. We find that the peak of the accreted mass attributed to the low energy flux model ($\log F = 10$, blue) is roughly co-located with a peak in the \TESS observations. This is in contrast with the the total $\rm{\dot{M}}$, which is somewhat misaligned with the peak in the \TESS light curve. \label{fig:TESS-HST-LDT_vs_time}}
\end{figure*}

GM Aur was unique in our sample in that we have additional U-band photometry and contemporaneous \HST STIS observations that we used to derive mass accretion rates. Here for the first time, we compare to mass accretion rates derived from our contemporaneous HST and U-band data.
Figure \ref{fig:TESS-HST-LDT_vs_time} shows the \TESS light curve and the \HST and LDT measurements of $\rm{\dot{M}}$. The top panel shows the \TESS light curve, while the middle panel shows the derived mass accretion rates from the U-band excess. $\rm{\dot{M}}$ measurements from the LDT U-band excess and those from \HST that are close in time are in excellent agreement given the uncertainties in measuring the mass accretion rate (e.g., 2019-12-07 UT and Epochs 10 \& 11). 

\begin{figure}
    \centering
    \includegraphics[width = 0.99\linewidth]{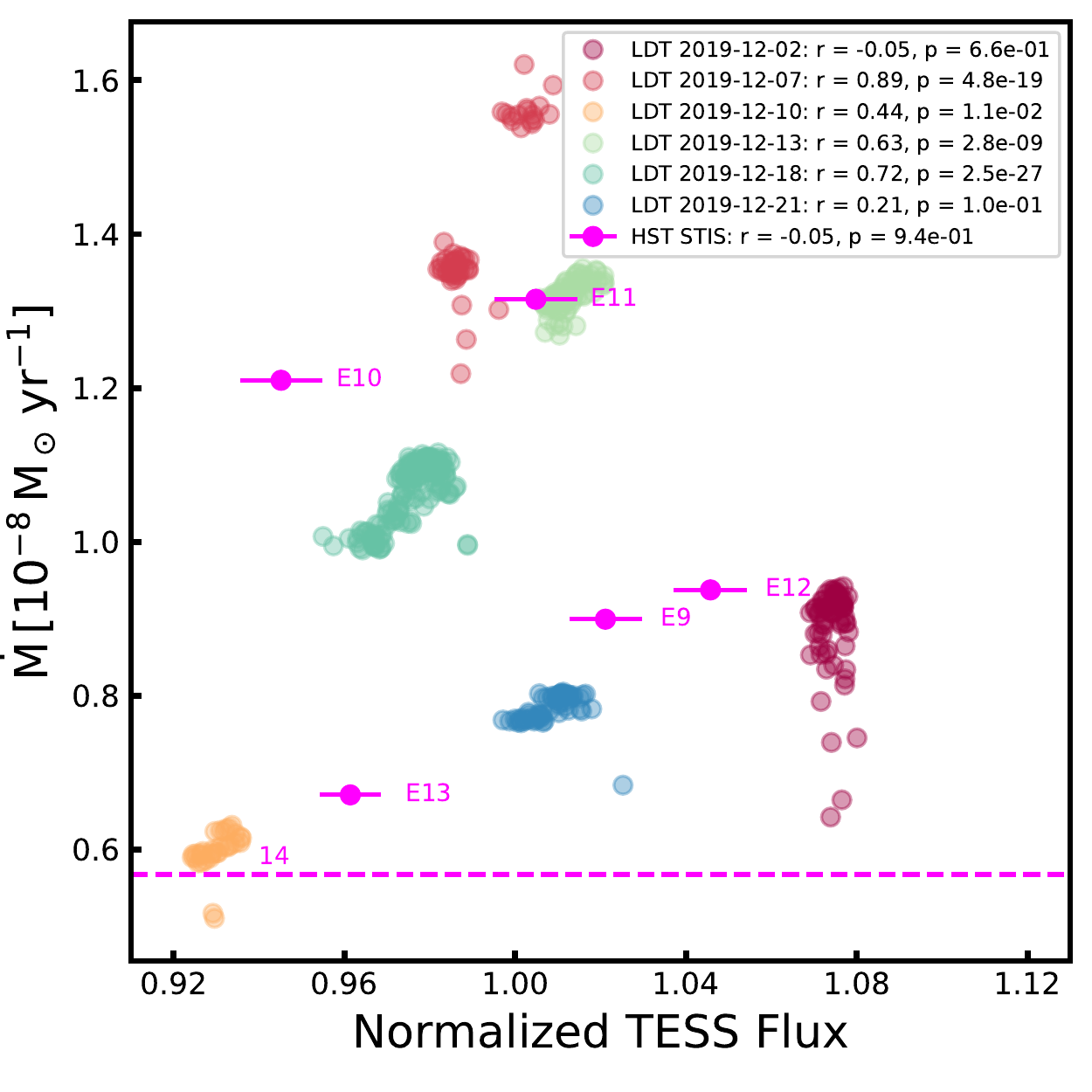}
    \caption{Normalized \TESS flux plotted against simultaneous $\rm{\dot{M}}$ measurements from HST and U-band photometry from the LDT for GM Aur. The different colors mark separate nights of LDT monitoring. The magenta bars associated with the \HST points span the range of the observed \TESS fluxes during that epoch. The dashed magenta line marks \HST Epoch 14, where the observation fell into the \TESS data downlink gap. The  Pearson correlation coefficients, $r$, and the associated $p$ values for each night of LDT observations and across all HST epochs are included in the legend.  \label{fig:GMAur-TESS-vs-mdot}}
\end{figure}

While correlations are present between the mass accretion rate and the observed \TESS flux during most of the individual nights of LDT observations, significant scatter is present when all of the observations are considered as a whole (see Figure \ref{fig:GMAur-TESS-vs-mdot}). The \textit{HST} observations mirror this and also display significant scatter across the multiple epochs of observations. We excluded the points in the \TESS light curve that were flagged as single-point outliers in this analysis (see Figure~\ref{fig:TESS_LC} and we did not identify any flares in the light curve of GM Aur from our by-eye inspection).

\begin{deluxetable*}{cccccc}
\centering
\tablecolumns{11}
\tablecaption{GM Aur Fitted Model Parameters \label{tab:gmaurfit}}
\tablehead{ Epoch & $\dot{M}$ & $f_{1E10}$ & $f_{1E11}$ & $f_{1E12}$ & $r_v$}
\startdata
9 & $0.901^{+0.009}_{-0.008}$ & $0.100^{+0.004}_{-0.004}$ & $0.0117^{+0.0003}_{-0.0003}$ & $0.000299^{+0.000012}_{-0.000012}$ & $0.549^{+0.009}_{-0.009}$ \\
10 & $1.210^{+0.010}_{-0.010}$ & $0.157^{+0.005}_{-0.005}$ & $0.0049^{+0.0005}_{-0.0005}$ & $0.001245^{+0.000021}_{-0.000021}$ & $0.553^{+0.009}_{-0.009}$ \\
11 & $1.310^{+0.010}_{-0.010}$ & $0.176^{+0.004}_{-0.004}$ & $0.00009^{+0.00009}_{-0.00003}$ & $0.001813^{+0.000016}_{-0.000016}$ & $0.665^{+0.010}_{-0.010}$ \\
12 & $0.938^{+0.009}_{-0.009}$ & $0.179^{+0.004}_{-0.004}$ & $0.00586^{+0.00026}_{-0.00026}$ & $0.000194^{+0.000010}_{-0.000010}$ & $0.665^{+0.010}_{-0.010}$ \\
13 & $0.666^{+0.007}_{-0.007}$ & $0.127^{+0.003}_{-0.003}$ & $0.00496^{+0.00017}_{-0.00018}$ & $0.000053^{+0.000006}_{-0.000005}$ & $0.477^{+0.009}_{-0.008}$ \\
14 & $0.568^{+0.007}_{-0.007}$ & $0.101^{+0.003}_{-0.003}$ & $0.00455^{+0.00018}_{-0.00018}$ & $0.000087^{+0.000007}_{-0.000007}$ & $0.442^{+0.008}_{-0.008}$ \\
\enddata
\tablecomments{Mass accretion rates, filling factors for low ($f_{1E10}$), medium ($f_{1E11}$), and high density ($f_{1E12}$) accretion columns, and veiling at V band, $r_v$, for GM Aur from the analysis of the \HST STIS observations with the accretion shock models presented in \citet{calvet98} and \citet{robinson19}. The subscripts following the filling factors refer to the kinetic energy flux of the accretion column (which is directly proportional to the column density) in units of $\rm{erg\, s^{-1} \, cm^{-2}}$. The value of each $f$ for each epoch is the fractional coverage of the visible stellar surface by shocks with that energy flux. Note that the uncertainties presented here are the $16^{th}$ and $84^{th}$ percentiles of the posteriors (roughly analogous to $1\sigma$) recovered from the MCMC analysis, and do not include systematic uncertainties (e.g., extinction and stellar parameters). \citet{robinson19} found that systematic uncertainties on the order of $\sim10\%$  are typically appropriate in derived parameters from this type of analysis. Results for Epochs 10, 11, 12, 13 and 14 were first presented in \citet{espaillat21}.}
\end{deluxetable*}

\begin{figure*}
    \centering
    \includegraphics[width = \linewidth]{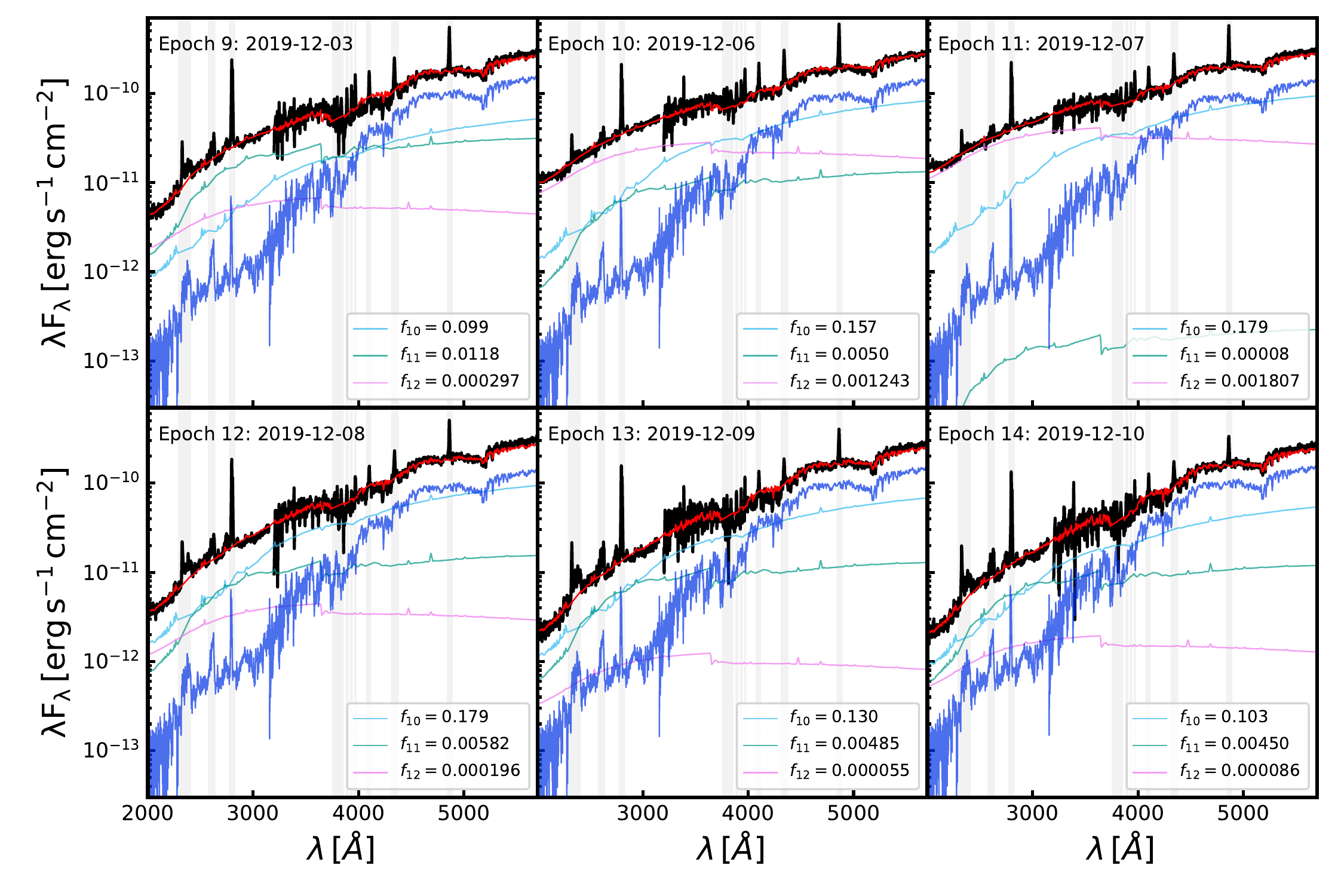}
    \caption{Model fits to the continuum emission of GM Aur during Epochs 9 - 14 as observed by STIS on \HST. The data are shown in black, and the total model is shown by the red line. The model components, $F = 10^{10}, \, 10^{11},$ and  $\, 10^{12} \, \mathrm{erg \, s^{-1} \, cm^{-2}}$ are shown in cyan, light blue, and pink respectively. The undisturbed photosphere model is displayed in dark blue. The total model shown here maximizes the posterior probability (i.e., the mode of the distribution). Note that this is slightly different from the median values reported in Table \ref{tab:gmaurfit}. Filling fractions for this model for each epoch are included in the legend. The light grey bars mark regions that were masked during the fitting procedure due to line emission. \label{fig:gmaur_fits}}
    
\end{figure*}

The contemporaneous NIR - NUV coverage of the STIS spectrograph allows us to break the degeneracy between accretion column density and accretion shock surface coverage. Table \ref{tab:gmaurfit} contains the fitted mass accretion rates, accretion shock filling factors, and veilings for \HST Epochs 9 - 14. The model with the maximum posterior probability is overlaid on the observations in Figure \ref{fig:gmaur_fits}. Individual contributions from low-, medium-, and high-density accretion shocks are shown separated from the total. The breakdown for Epochs 10, 11, 12, 13, and 14 were presented by \citet{espaillat21}. Here were present the distribution for Epoch 9 for the first time. We find that the results for this epoch are are in good agreement with previous measurements of GM Aur in a intermediate accretion state, with a shock surface coverage fraction of about 11\% and a mass accretion rate of $0.9 \times 10^{-8}\rm{M_\odot \, yr^{-1}}$.

\subsection{Connecting \TESS to accretion on different timescales \label{ss:time_lags}}

\subsubsection{Sub-exposure Time-tag Analysis \label{ss:gmaur_timetag}}
\begin{figure}
    \centering
    \includegraphics[width = 0.99\linewidth]{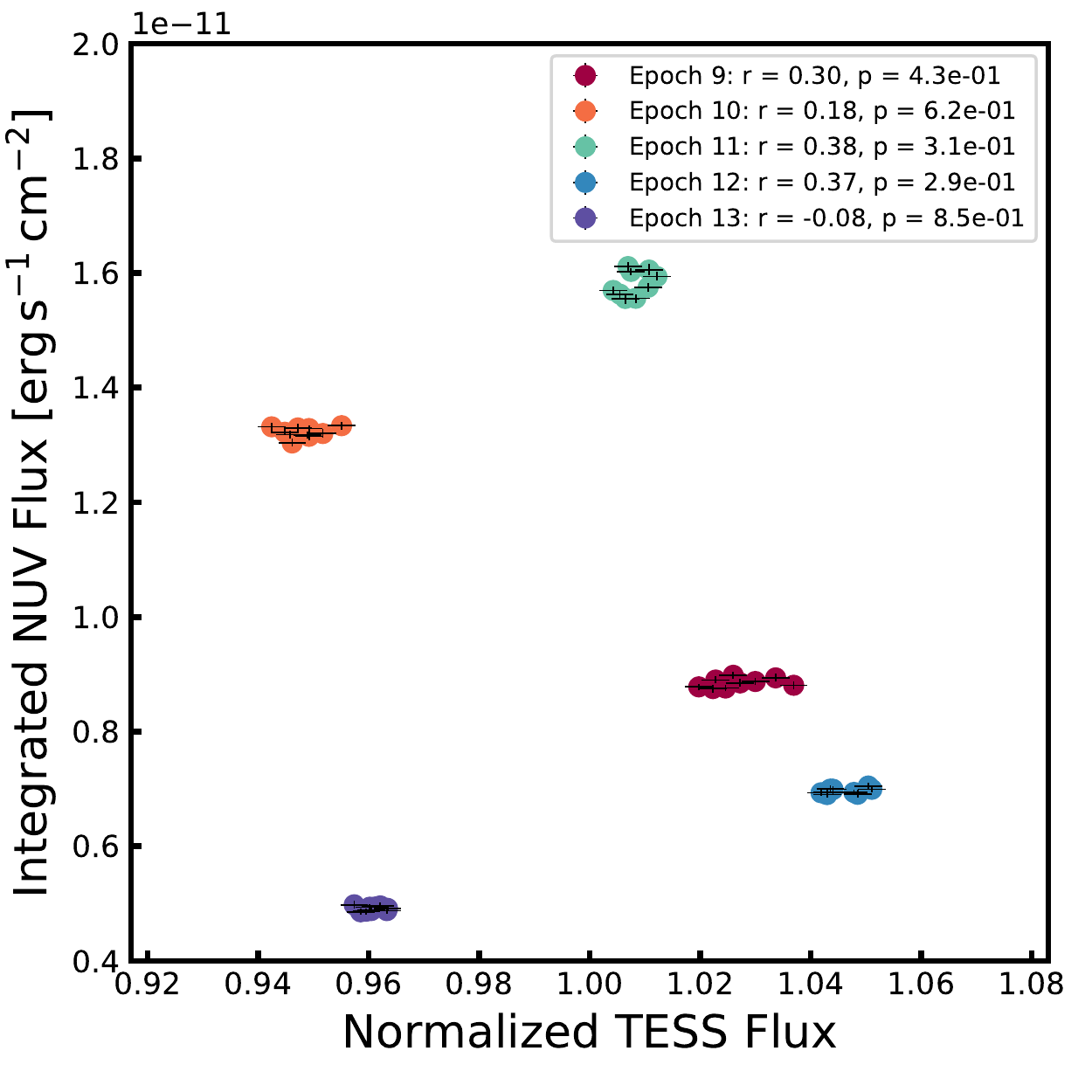}
    \caption{Integrated NUV flux from 120s sub-exposures of the NUV \HST STIS observations of GM Aur plotted against the simultaneous \TESS observations. The Pearson correlation coefficient $r$ and its $p$ value for each visit are included in the legend. We find hints of weak positive trends during four out of five epochs, but note that these trends are not strongly significant based on their associated $p$ values. We find no correlation between multiple epochs. \label{fig:time_tag}}  
\end{figure}

The STIS NUV-MAMA detector records photon arrival timestamps while in the time-tag mode which can be used to break long, single exposures into multiple, shorter sub-exposures. By breaking our longer NUV GM Aur exposures into 120 s chunks and adding a delay to the first sub-exposure, we obtain exactly simultaneous NUV-\TESS observations which we use to search for correlations on minute-timescales. Figure \ref{fig:time_tag} shows the integrated NUV flux between $1700 - 3100$ \angstrom~plotted against the simultaneous \TESS observations. 

The existence of a trend between \HST and \TESS within individual \HST visits is tested by calculating the Pearson correlation coefficient, $r$ and the associated $p$ value for each visit. We find weak, positive trends between \TESS and \HST within most individual exposures, but note that the reported $p$ values shows that the trends are not strongly statistically significant. We attribute this in part to random uncertainties which are comparable to the measured differences in flux between sub-exposures and the small number of sub-exposures (typically $\sim10$) obtained during each \HST visit. Values of $p$ and $r$ for each visit are reported in Figure \ref{fig:time_tag}. Ultimately, the observational uncertainty in our measurements prevents us from firmly establishing or ruling out a strong relationship between the emission in the NUV and in the \TESS passband on these short timescales, but hints of such a correlation may be present.

\subsubsection{Identifying time-lags on hour-to-day timescales \label{sss:cross_ref_lag}}
The models of \citet{calvet98} find that the excess emission produced by higher-density accretion columns tend to peak at shorter wavelengths when compared to that of lower-density columns. Previous work has found that some objects require both high and low density components to reproduce their observed emission \citep[e.g.,][]{ingleby13}. Using these multi-column models to fit Epochs 10, 11, 12, 13 and 14 of the \HST STIS spectra of GM Aur, \citet{espaillat21} identified a one day lag between the UV emission coming from the high-density columns and the optical emission from the low-densities columns (see Figure~\ref{fig:TESS-HST-LDT_vs_time}). This was presented as evidence for longitudinal density stratification in the accretion column. The peaks of the filling factors for the low density shocks are also well-aligned with the peaks in the simultaneous \TESS light curve, which is reasonable because the \TESS bandpass is centered at red/infrared wavelengths where the emission from the low-density accretion columns is higher \citep[e.g.,][]{calvet98}. This may also explain some of the scatter in the slopes between the excess \TESS luminosity and the values of $\rm{\dot{M}}$ measured using the excess in the U-band between objects (Figure~\ref{fig:TESS-vs-Lacc}), since the U bandpass is more sensitive to the higher density accretion accretion column energy fluxes than the \TESS bandpass. For a visualization of this longitudinal density structure, see Figures 3 and 4 in \citet{espaillat21} which present a model of the magnetosphere and a surface map of the accretion column density for a CTTS with such gradients.

We attempt to identify similar lags by cross-correlating our \TESS data with our UBVRI photometry using the Python-based package \textit{Stingray} \citep{stingray}. We make the assumption that our missing UBVRI data can be considered Missing Completely At Random (MCAR), and impute the mean value for the missing values. Note that the UBVRI data is very sparse (only 6 observations compared to the $\sim 18000$ \TESS points in each light curve), so the lags identified by this method may not be unique. We exclude regions in the \TESS light curve that we identified as having stellar flares or as single-point outliers (see Figure~\ref{fig:TESS_LC}), but note that their inclusion does not significantly change our results. We smoothed our cross-correlation results using a 3rd-order Savitzky-Golay filter \citep{savitzky64}. We then identified peaks in the smoothed function by applying a threshold that the minimum peak value must be greater than 0.2 times the maximum peak and separated by least least $\sim~1.5$ days. This process of smoothing and thresholding reduces the number of detected peaks to only those that are most significant. 

\begin{figure*}
    \centering
    \includegraphics[width = \linewidth]{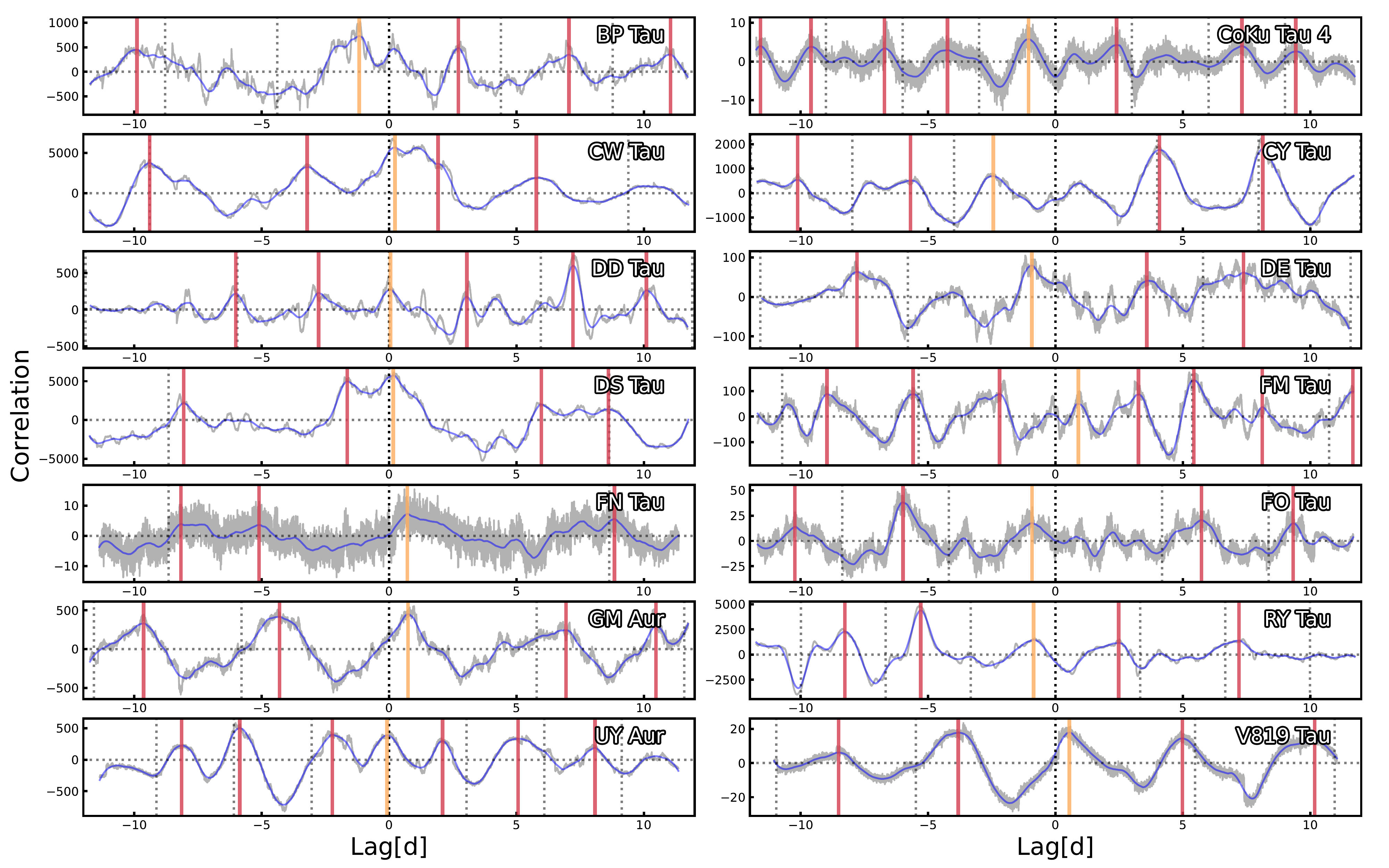}
    \caption{Cross-correlation between the high-cadence \TESS light curves and the sparse, ground-based, U-band light curves for all 14 objects. The grey line is the raw correlation function, while the blue line shows that function smoothed using a Savitzky-Golay filter. Identified peaks are marked as red and gold vertical lines, where the gold line marks the peak with the smallest absolute lag. The vertical dashed lines mark the rotation period of the star in lag-space. This process was also repeated for each of the other bands (BVRI).}
    \label{fig:U_cross_corr}
\end{figure*}

\begin{figure*}
    \centering
    \includegraphics[width = \linewidth]{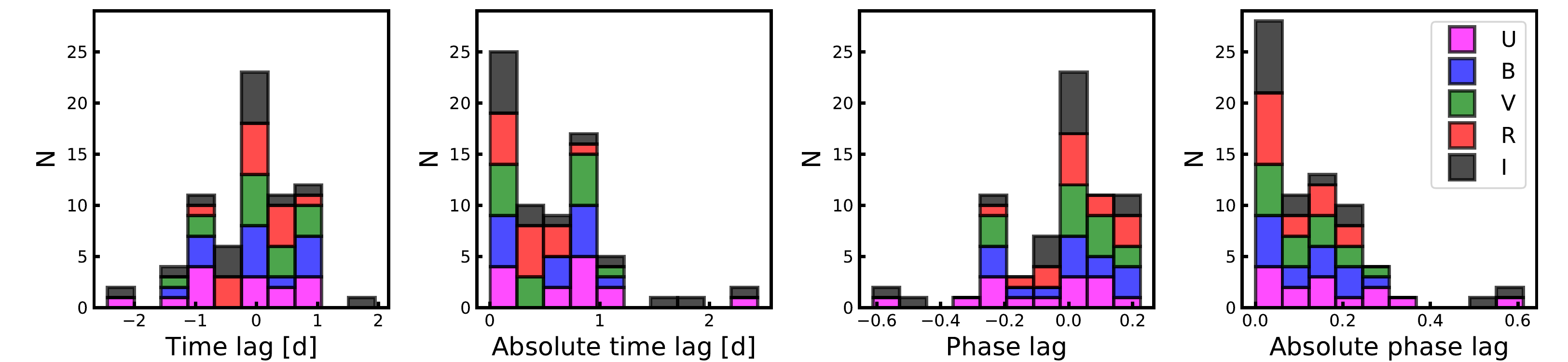}
    \caption{Histograms of the time lag, absolute time lag, phase lag, and absolute phase lag for the peak in the cross-correlation function regarded as significant by our criteria whose absolute time lag is closest to 0 lag (see the gold vertical lines in Figure~\ref{fig:U_cross_corr}). We found that the median absolute time lag and the median absolute phase lag tend to increase at short wavelengths, which may indicate longitudinal density stratification of the accretion shock. }
    \label{fig:lag_hist}
\end{figure*}

We show the results from this analysis for our U-band photometry in Figure.~\ref{fig:U_cross_corr}. Histograms of the identified peaks for each object with absolute value closest to 0 d are shown in Figure.~\ref{fig:lag_hist}. The respective median absolute lags for our UBVRI photometry are 0.802, 0.673, 0.443, 0.352, 0.365 d, while the respective rotational-phase lags are 0.145, 0.122, 0.085, 0.060, 0.064. This anticorrelation between wavelength and median lag supports the idea that longitudinal stratification of accretion column density may be common for CTTS. The $\sim 1$ d lag between our U-band photometry and the \TESS light curve found here for GM Aur is in good agreement with the lag between high- and low-density accretion columns identified by \citet{espaillat21}. Because of our very coarse sampling, it is not surprising that we do not find a median lag of zero between our \TESS observations and our I-band photometry.

\section{Discussion \label{s:discussion}}

\subsection{\TESS as a tracer of accretion \label{ss:tess_and_mdot}}
We identify positive correlations between $\rm{L_{acc}}$ measured through U-band measurements and the \TESS flux scaled using our I-band photometry (see Figure~\ref{fig:TESS-vs-Lacc}). Furthermore, we do not see strong positive correlations between objects with variability classifications not associated with accretion within our sample. We also identify correlations between \TESS and $\rm{\dot{M}}$ during the hour-long monitoring sessions of GM Aur.
However, when we examine the slopes of the linear regressions, the idea of a single, global relationship between \TESS and $\rm{L_{acc}}$ (analogous to that of $L_{acc}$ and $L_{U}$) begins to break down. This is seen both across our entire sample (see the bottom right panel of \ref{fig:TESS-vs-Lacc}) and between separate nights of observations for GM Aur (see Figure~\ref{fig:GMAur-TESS-vs-mdot}).  Likewise, on even shorter scales we find different slopes between \TESS flux and integrated NUV flux (a proxy for the mass accretion rate) from the time-tag analysis with the \HST data (see Figure \ref{fig:time_tag} and \S\ref{ss:gmaur_timetag}, but note that the correlations are very weak in those observations. The results of the decomposition of the accretion columns into individual components presented in Table~\ref{tab:gmaurfit} and Figure~\ref{fig:gmaur_fits} as well as the time-lags shown in \S~\ref{ss:time_lags} also demonstrate that the \TESS bandpass misses some of the information about accretion encoded in the shorter wavelengths. 

In short, we suggest that any global relationship between observed \TESS flux and the mass accretion rate would be unreliable, but note that \TESS flux does trace facets of the accretion behavior for individual objects particularly on shorter timescales. We suggest that some of the discrepancies between the mass accretion rate and the observed \TESS flux may arise because of longitudinal stratification of accretion column densities \citep[see Section~\ref{ss:time_lags}, also][]{espaillat21}. \TESS is primarily sensitive to low-density columns. The \TESS bandpasses and the inferred $\dot{M}$ values are also sensitive to other sources of astrophysical variability such as disk obscuration or star spots \citep[which can be large for TTS][]{bradshaw14} which could introduce similar effects. We suggest that this non-accretion-related variability may in part lead to the observed differences in relationships between the two quantities for individual sources \cite[analogous to the scatter in color slopes for the sample presented in][]{venuti15}.

We searched for trends between the color slopes shown in Figure~\ref{fig:cmd} and our measured values of Q and M, but we did not find obvious trends (likewise with our by-eye classifications). We also did not identify strong correlations between our mean mass accretion rates and Q and M (even when excluding the two dippers in our sample, CY Tau and RY Tau). While our sample size is limited, both results are in agreement with the wide spread of slopes and accretion rates that have been observed in similar studies with larger samples of CTTS \citep[e.g.,][]{venuti15, sousa16}.

\subsection{Connecting light curve variability metrics to literature data \label{ss:literature}}
\begin{figure*}
    \centering
    \includegraphics[width = 0.7\linewidth]{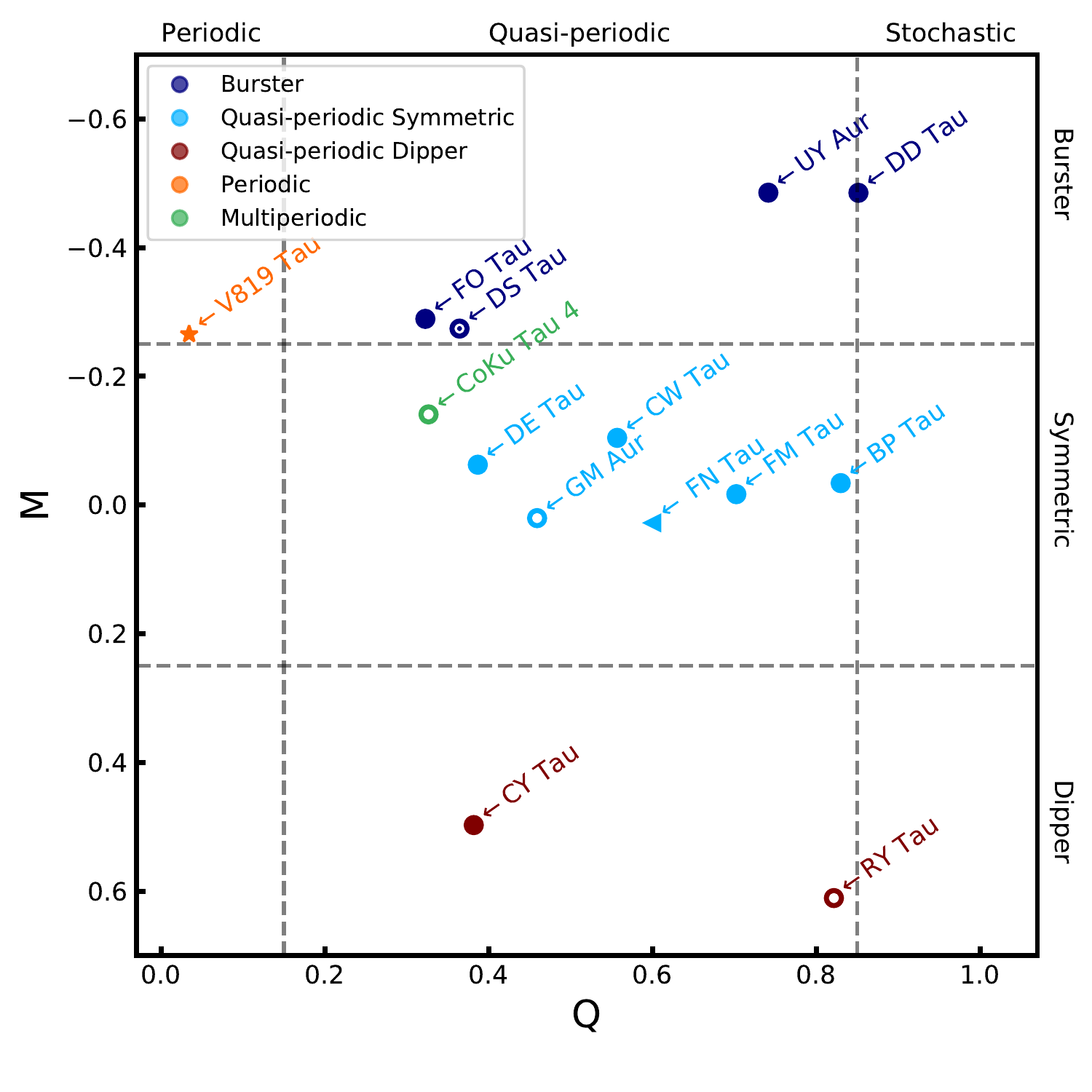}
    \caption{Bivariate plot of the statistical metrics Q (periodicity) and M (symmetry) for the \textit{TESS} light curves \citep[see][for more details on Q and M]{cody14}. Filled circles are full disks, empty circles are transitional disks, and concentric circles are pre-transitional disks. V819 Tau is a debris disk (marked with a star), and FN Tau is a full disk with an upper limit on Q. We adopt the boundaries between variability classes from \citet{cody18}, which are marked by the dashed lines. }
    \label{fig:QM}
\end{figure*}

The nearby, well-studied nature of our sample allows us to make connections that have not been possible in previous space-based monitoring campaigns \citep[e.g.,][]{cody14, cody18}. Here we compiled ancillary literature data when available. Figure~\ref{fig:QM} plots the 14 objects in Q-M space and indicates their variability classification as well as their disk morphology as measured from IR/mm observations (see Table~\ref{tab:stellar_params}). While we note that our sample is small, we can make a few observations. None of the disks that are classified as B are transitional or debris disks, which may have limited material in the innermost regions compared to full or pre-transitional disks. Of the three transitional disks in our sample, Coku Tau 4 is classified as MP, GM Aur as QPS, and RY Tau as QPD. 

\begin{figure*}
    \centering
    \includegraphics[width = \linewidth]{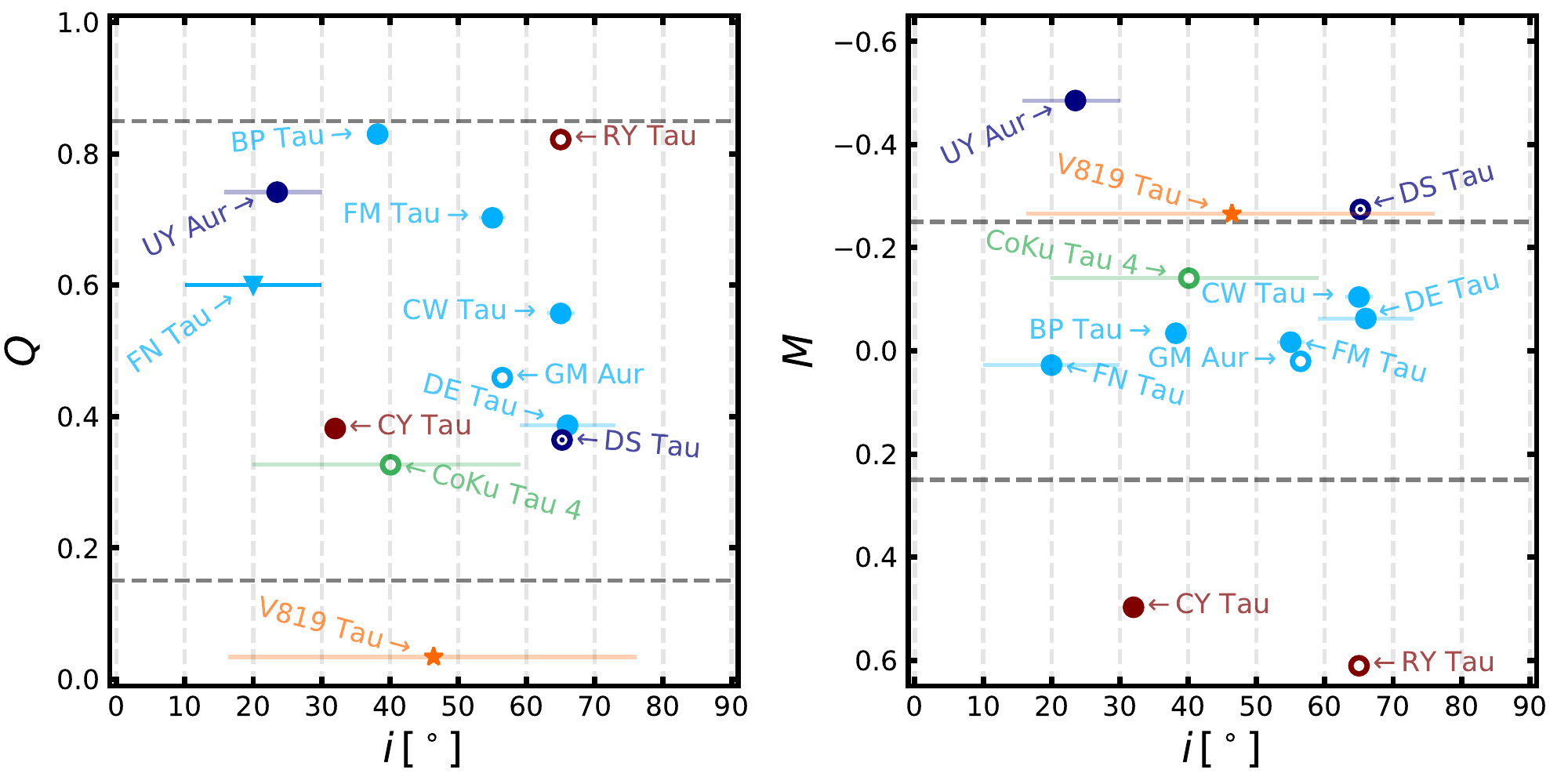}
    \caption{\textit{Left}: \TESS light curve periodicity (Q) plotted against disk inclination ($i$). For objects that have light curve variability associated with accretion (QPS in light blue, B in purple) we see hints of a weak inverse correlation between inclination and Q. \textit{Right}: Light curve symmetry (M) against disk $i$. A clear pattern between M and $i$ for the accreting objects is difficult to identify. CY Tau is an interesting outlier as a nearly face-on object displaying a dipping light curve.}
    \label{fig:QM_i}
\end{figure*}

Disk inclinations can be measured through resolved mm studies of disks. The right panels of Figure~\ref{fig:QM_i} show disk inclinations reported from several studies in the literature (see Table~\ref{tab:stellar_params}) and the Q and M metrics measured from our \TESS data. As noted by previous authors \citep[e.g.,][]{cody18}, distinct trends between these variability classifications are difficult to identify. However, we find a weak inverse trend between disk inclination and Q for objects with variability classifications associated with accretion (QPS and B). We find little evidence for a strong correlation between M and $i$ for these objects. Both are in good agreement with the results of \citet{robinson21}, which showed that for accreting objects, Q varied across the entire range of inclinations while M was only a strong function of $i$ when viewed nearly edge-on ($i \gtrsim 70^\circ$).

One interesting object to note is CY Tau, which we classify as quasi-periodic dipper. The dips in this object are narrow and deep, and occur at a period similar to the apparent rotation period of the star. This may suggest an disk warp induced by the stellar magnetic field in the inner regions of the disk \citep[perhaps analogous to the geometry of V354 Mon presented by][]{schneider18}. However, the disk inclination as traced by mm observations is unambiguously close to face-on \citep[$i = 32^{+1}_{-1}{}^\circ$,][]{simon17}. 
This suggests that the stellar inclination and/or the inner disk may be strongly misaligned with mm measurements of the disk inclination. This is something that has been observed for other dippers, including the prototypical example AA Tau \citep{loomis17}. Misalignment of inner disks has been suggested to be a possible indicator of companions \citep{price18} or planetary mass objects \citep{zhu18}.

\subsubsection{Comparing BP Tau to simulations}
The simulations of \citet{robinson21} made predictions about what system parameters are important for setting the morphology of light curves for young stars with rotation-axis-aligned magnetic field topologies. The parameters tested in that work include the stellar mass, the co-rotation radius, the ratio between octupole and dipole magnetic field coefficients, inclination, and the turbulence mach number in the inner disk. Within our sample, BP Tau is the only non-periodic object within our sample that has a published measurement of the magnetic field topology \citep{donati08b}. Its magnetic field consists of a 1.2 kG dipole and 1.6 kG octupole both tilted slightly ($\sim 10^\circ$) with respect to the stellar rotation axis with a relatively small toroidal component. Coincidentally, the ratio of the octupole to dipole field strengths and the other system parameters of BP Tau are relatively similar to one of simulations of \citet{robinson21} (the model in question assumes a turbulent mach number of 0.1 in the inner disk). We find that our measured Q and M values are consistent with predictions from that simulation within their model uncertainties. While potentially interesting, magnetic topology measurements exist for only one of our accreting targets and we caution against further conjecture at this point. Magnetic topology observations do exist for V819 Tau \citet{donati15}, but that source is a WTTS with a magnetic field composed primarily of a dipole that is significantly tilted with respect to the rotation axis ($\sim 30^\circ$), so the simulations of \citet{robinson21} are likely less applicable in that case. 

\subsection{Comparisons to previous GM Aur \texorpdfstring{$\dot{M}$}{mdot} measurements}
The values of the mass accretion rates of GM Aur measured from the \HST data and the LDT are generally consistent with the previously observed range during Epochs 1-8, which spans from $0.6 - 2.0 \times 10^{-8} \, \mathrm{ M_\odot \, yr^{-1}}$ \citep{ingleby15, robinson19}. The observed range of NUV emission for Epochs 1-8 is shown in Figure \ref{fig:gmaur_spectra}. When compared to the other previously analyzed epochs, the inferred distribution of accretion column densities during Epoch 9 is typical. The range of veiling values estimated from the $s$ parameter for Epochs 9-14 is comparable to that of Epochs 1-8. 

While the reported mass accretion rates from this work span roughly a factor of two, the model of azimuthal density structure across the accretion column from \citet{espaillat21} suggests that this may not be an accurate measurement of the global mass accretion rate but instead primarily caused by changes in the viewing geometry. In contrast, \citet{robinson19} identified an accretion burst during Epoch 7 that remains remarkable. The peak accretion rate during that event was $\sim20\%$ higher than the peak accretion measured from our LDT photometry, and $\sim100\%$ higher than the mean accretion rate. Critically, the inferred distribution of accretion columns during this epoch were also distinct when compared with all other 13 epochs, with the largest contribution to the total mass accretion rate being from high-density columns. This event apppears to be a true increase in the global accretion rate facilitated by a a particularly dense accretion column, which supports the idea that that large density gradients may be present in the inner disk. This accretion burst with the additional context of rotational modulation of complex azimuthal density structures provided by Epochs 9 - 14 demonstrates the strength of multi-wavelength monitoring for better understanding the structure and volatility of the inner disk.

\section{Summary \label{s:summary}}
We observed 14 objects using the short-cadence mode on \TESS for 26 d. During that time, we also obtained contemporaneous UBVRI photometry with the LDT on six nights for each object and monitored GM Aur over hour-timescales in U. Additionally, six epochs of \HST NUV-NIR STIS data were obtained simultaneously for GM Aur during the \TESS observations. With the \TESS data, we estimate rotational periods and determined empirical variability classifications through the use of previously developed statistical metrics and by-eye checks. With the LDT data, we measured mass accretion rates through the measured U-band excess, searched for correlations between $\rm{L_{acc}}$ and readily available \TESS data, and identified time-lags with our simulatenous \TESS data. With the \HST data, we present new analysis for one epoch \citep[the other five epochs were presented in][]{espaillat21}, compare the measured mass accretion rate to those derived from our U-band photometry, and searched for connections to \TESS on minute-timescales using time-tagged observations. Our primary findings are as follows:

\begin{enumerate}
    \item Some of the variability present in the \TESS light curves can be linked to accretion. 
    \item However, a single global relationship between \TESS flux and $\rm{L_{acc}}$ is not readily apparent. This is demonstrated by the differences in slope between different objects, and even individual objects on different nights.
    \item Our measured accretion rates for GM Aur from our U-band photometry agree very well with those derived from detailed accretion shock modeling of contemporaneous NUV-NIR measurements from \textit{HST}. 
    \item We found some additional evidence for longitudinal density stratification of accretion columns by identifying time lags within our dataset that increase at shorter wavelengths.
    \item We identified CY Tau as a face-on dipper, which may suggest the presence of a misaligned inner disk and/or magnetic field.  
    \end{enumerate}
We conclude that \TESS light curves do trace some of the accretion behavior on shorter timescales, but are not reliable tracers of the total accretion rate, especially over longer timescales. These results highlight the importance of obtaining simultaneous, multi-wavelength observations when interpreting \TESS light curves. This will become increasingly important as \TESS light curves continue to become available for CTTS. 

\section{Acknowledgements}
This work was supported by {\it HST} grant GO-16010, NASA grant 80NSSC19K1712, and NSF Career grant AST-1455042.
Support for program \#16010 was provided by NASA through a grant from the Space Telescope Science Institute, which is operated by the Association of Universities for Research in Astronomy, Inc., under NASA contract NAS5-26555.

The authors thank the members of Dr. Connor Robinson's Ph.D. defense committee, Dr. Philip Muirhead, Dr. Merav Opher, Dr. Daniel Clemens, and Dr. James Owens, whose comments greatly improved this document during the preparation of his Dissertation. We also thank Dr. Nuria Calvet, and our anonymous reviewer whose insightful comments improved this work. 

This paper includes data collected by the TESS mission, which are publicly available from the Mikulski Archive for Space Telescopes (MAST). Funding for the TESS mission is provided by NASA's Science Mission directorate.

Based on observations made with the NASA/ESA Hubble Space Telescope, obtained at the Space Telescope Science Institute, which is operated by the Association of Universities for Research in Astronomy, Inc., under NASA contract NAS5-26555. These observations are associated with program \#16010.

These results made use of the Lowell Discovery Telescope (LDT) at Lowell Observatory. Lowell is a private, non-profit institution dedicated to astrophysical research and public appreciation of astronomy and operates the LDT in partnership with Boston University, the University of Maryland, the University of Toledo, Northern Arizona University and Yale University. The Large Monolithic Imager was built by Lowell Observatory using funds provided by the National Science Foundation (AST-1005313).

This work has made use of data from the European Space Agency (ESA) mission
{\it Gaia} (\url{https://www.cosmos.esa.int/gaia}), processed by the {\it Gaia}
Data Processing and Analysis Consortium (DPAC,
\url{https://www.cosmos.esa.int/web/gaia/dpac/consortium}). Funding for the DPAC
has been provided by national institutions, in particular the institutions
participating in the {\it Gaia} Multilateral Agreement.

This research made use of Photutils, an Astropy package for
detection and photometry of astronomical sources \citep{bradley20}. 

\section{Appendix: Remarks on Individual \TESS Light Curves \label{s:indiv}}

Here, we discuss individual objects with regards to their observed variability classifications and stellar activity. Derived statistical metrics Q and M, periods, and empirical variability classifications can be found in Table~\ref{tab:QMT}. This section refers to events occurring in terms of days since MBJD 58815 in the TDB time scale (see Figure~\ref{fig:TESS_LC}).

\textbf{BP Tau}
Using Q and M, BP Tau is classified as a QPS object. Periodic behavior in the light curve is not immediately obvious by-eye. Some features that resemble both dips and accretion bursts are present. A bright flare was observed near 3.5d. Typical changes in brightness are on the order of $10\%$. 

\textbf{Coku Tau 4:}
We distinctly see periodic behavior with two frequencies and thus classify CoKu Tau 4 as an MP object rather than the QPS that Q and M would suggest. The amplitudes of observed changes for this object are small, on the order of $1\%$ and are likely from rotational modulation of stable hot/cold spots. We do not see obvious evidence for flaring in this object.
A Lomb-Scargle Periodogram  \citep{lomb76, scargle82} was used to identify both periods of the signals present in the light curve.

\textbf{CW Tau:}
This object exhibits very large changes in flux (up to $50\%$) with limited obvious periodicity, hence its classification as a QPS. Most of the changes resemble accretion events, but some events may be dips. We do not see evidence of flaring, but note that flares may be masked by the strong variability. 

\textbf{CY Tau:}
This object exhibits periodicity and shows evidence for both accretion bursts and dips, resulting in a classification of QPD. As discussed in \S~\ref{ss:literature}, this object has an inclination of \citep[$i = 32^{+1}_{-1}{}^\circ$,][]{simon17}, which may suggest the presence of a misaligned inner disk and/or stellar inclination.  A single flare was identified near 11.5d. 

\textbf{DD Tau:}
Some degree of periodicity is visible for this object. Most of the variability resembles accretion bursts, resulting in a B classification. Two flares were identified near 1.2d and 24.8d. 

\textbf{DE Tau:}
This object shows a periodic signal with accretion bursts and was classified as a QPS object. Two flares are present near 2.0d and 24.0d. 

\textbf{DS Tau:}
A possible period can be identified for this object, but it is somewhat masked by changes in the accretion rate. This object displays large changes in flux (up to $50\%$) and was assigned a B classification. 

\textbf{FM Tau:}
Clear periodicity is difficult to identify for this object. The light curve is symmetric with both dimming and brightening events, resulting in a QPS classification. Two flares were observed, near 6.8d and 21.0d. Relative changes for this object are moderate (on the order of $25\%$). 

\textbf{FN Tau:}
While the changes in this object are quite small (on the order of $1\%$), a periodic signal is identifiable. We found that the \TESS flux uncertainties from the SPOC pipeline were slightly overestimated compared to the observed scatter, causing the algorithm used to measure Q to fail.  To account for this, we report an upper limit on Q for this object and note that both P and QPS are possible variability classifications. Two potential flares were identified near 6.0d and 21.7d. 

\textbf{FO Tau:}
A period is identifiable and accretion bursts are present in the \TESS light curve, resulting in a B classification via Q and M. A single flare was identified near 25.0d. No quasi-periodicity is detected above a significance of 6 standard deviations. 

\textbf{GM Aur \label{ss:gmaur}}
GM Aur is unique within the sample because of the NUV - NIR \HST STIS data obtained simultaneously with the \TESS observations and the additional U-band monitoring sessions. More analysis on GM Aur is presented in \citet{espaillat21}. 
 From the \TESS light curve, a period can be identified. On top of the periodic behavior, we see moderate increases, likely associated with changes in the accretion rate, resulting in a classification of QPS. No obvious flares were identified for this object. 

\textbf{RY Tau:}
Deep dips are visible during the first half of the \TESS observations which contributes to the classification of QPD. During the second half, almost no period is visible.  RY Tau has been previously observed to show dipping behavior thought to be associated with variations in the upper layers of the inner disk \citep[][]{davies20} which matches the observed behavior here. No flares were identified for this object. 

\textbf{UY Aur:}
A clear periodic signal can be identified, punctuated by flares and accretion events, resulting in a classification of B. We identified 5 events that resemble flares in this object, one of which appears to be at the beginning of a major accretion event. 

\textbf{V819 Tau:}
V819 Tau is a very periodic source with many flares. We identified 10 events that resemble the characteristic rapid rise and exponential decay of flares. We see very limited variability that resembles dipping/accretion. 

\bibliography{biblio}
 
\end{document}